\documentclass[aps,prb,10pt,
twocolumn,
showpacs,superscriptaddress]{revtex4-2}
\usepackage{amsmath, amssymb, amsfonts}
\usepackage{xcolor}
\usepackage{graphicx}
\usepackage{bm}
\usepackage[squaren]{SIunits}
\usepackage{multirow}
\usepackage[colorlinks=true,allcolors=blue]{hyperref}

\begin{document}

\title{
Novel implementations for reservoir computing -- from spin to charge }

\author{Karin Everschor-Sitte}
\affiliation{Faculty of Physics and Center for Nanointegration Duisburg-Essen (CENIDE), University of Duisburg-Essen, 47057 Duisburg, Germany}

\author{Atreya Majumdar}
\affiliation{Faculty of Physics and Center for Nanointegration Duisburg-Essen (CENIDE), University of Duisburg-Essen, 47057 Duisburg, Germany}

\author{Katharina Wolk}
\affiliation{Department of Materials Science and Engineering, Norwegian University of Science and Technology (NTNU), Trondheim 7034, Norway}

\author{Dennis Meier}
\affiliation{Department of Materials Science and Engineering, Norwegian University of Science and Technology (NTNU), Trondheim 7034, Norway}
\affiliation{Center for Quantum Spintronics, Norwegian University of Science and Technology (NTNU), Trondheim 7034, Norway}

\date{\today}

\begin{abstract}
Topological textures in magnetic and electric materials are considered to be promising candidates for next-generation information technology and unconventional computing. Here, we discuss how the physical properties of topological nanoscale systems, such as skyrmions and domain walls, can be leveraged for reservoir computing, translating non-linear problems into linearly solvable ones. In addition to the necessary requirements of physical reservoirs, the topological textures give new opportunities for the downscaling of devices, enhanced complexity, and versatile input and readout options. Our perspective article presents topological magnetic and electric defects as an intriguing platform for non-linear signal conversion, giving a new dimension to reservoir computing and in-materio computing in general.
\end{abstract}

\pacs{}

\maketitle


\section{Introduction}
\label{sec:intro}
Analog computing schemes rely on the continuous responses of physical, biological, and chemical systems for data storage and information processing. One major advantage of analog computing is the possibility of \textit{in-memory} processing without the need to convert and transfer data as required in von Neumann architectures~\cite{Ielmini2018}.
As a consequence, certain operations can be performed much faster than with digital computers and at reduced energy costs, offering great potential for the design of next-generation information technology~\cite{Indiveri2011}. Image recognition, classification, and non-linear signal prediction are examples of fields where brain-inspired unconventional computing methods can outperform digital systems with respect to speed and energy consumption. The concept of unconventional computing, however, is not new, and the quest for suitable material systems is well on the way~\cite{Finocchio2021, Finocchio2023}. 

The discovery of functional topological spin structures, i.e., particle-like magnetic objects, has given an intriguing new twist to the field by introducing a novel form of physical implementation that can be utilized for computation~\cite{Lee2023a}. Among the most intensively researched topological spin textures are magnetic skyrmions, which are stable whirls formed by spins that can be efficiently excited or moved by electrical currents~\cite{Fert2017, Nagaosa2013}. Such skyrmions are being used, for instance, to design neuron devices and artificial synapses, emulating the behavior of biological systems~\cite{Back2020, Vedmedenko2020, Lee2023a}. Other topological spin textures that currently attract attention include anti-skyrmions, hopfions, and dislocations, offering fertile ground for the design of unconventional computing schemes~\cite{Tang2021, Wang2019, Kent2021, Azhar2022, Stepanova2021}.

In a more recent development, topologically protected ami arrangements of electric dipoles in ferroelectrics are explored for analog computing~\cite{Mcconville2020, Rieck2023, Meier2022, Sharma2022, Wang2022}. The research gained substantial momentum due to the discovery of ferroelectrics compatible with complementary metal–oxide–semiconductor (CMOS) processes, which is essential for device fabrication~\cite{Schroeder2013}. In close analogy to magnetic materials, electric skyrmions, topologically non-trivial domain walls, and other exotic electric dipole textures have been reported~\cite{Govinden2023, Nataf2020, Das2019} and related possibilities for device applications are being discussed~\cite{Focusissue}. In comparison to magnetic materials, however, the potential of topological dipole textures in ferroelectrics for analog computing schemes is much less explored.

The opportunities that arise from topological magnetic and electric textures, and the close connection between their topologies, motivate this perspective article. Our focus is on innovative approaches for reservoir computing enabled by selected topological nanoscale textures, and the potential they offer for \textit{in-materio} computing and possible future devices.

\section{Reservoir computing}

\begin{figure}[tb]
\includegraphics[width=0.75\columnwidth]{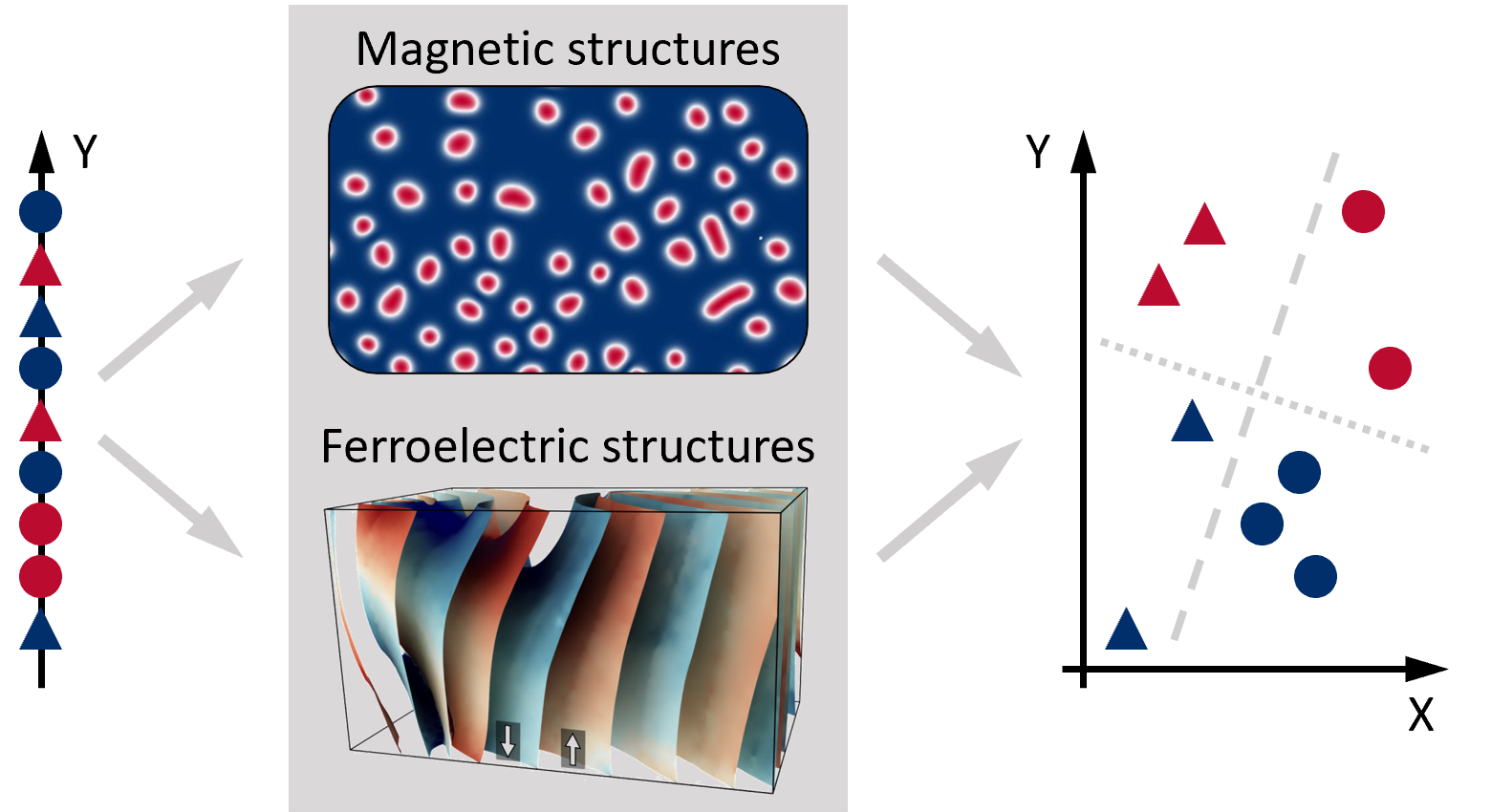}
\caption{Physical reservoir computing scheme exemplified by a magnetic or ferroelectric reservoir (center). The reservoir transforms a nonlinear problem (left) into a linear task in higher dimensions (right). The same reservoir is capable of addressing multiple problems; this is achieved by re-training only the last readout layer. This capability is illustrated here by two distinct lines: one differentiates by color (red and blue), whereas the other differentiates by shape (circles and triangles). Images are modified versions from data published in Ref.~\cite{Grollier2020, Roede2022}.}
\label{fig:RCscheme}
\end{figure}

Reservoir computing is a type of unconventional computing scheme that excels at the recognition and prediction of spatio-temporal events~\cite{Lukosevicius2012, Jaeger2001}. In this computing paradigm, the eponymous reservoir is a system that maps the input into a higher-dimensional space where a non-linear problem transforms into a linearly (or, more generally, simply) solvable one. Figure~\ref{fig:RCscheme} illustrates this process for a one-dimensional arrangement of symbols, specifically triangles and circles. The reservoir projects these symbols into a two-dimensional space, allowing them to be separated by a straight line based on their shape, color, or both. 

In reservoir computing, the input is processed by the reservoir and then passed through a trained readout layer to produce the final output. This type of computing only needs training for this final linear layer, making it more computationally efficient than deep neural networks, which require training at each layer. For large datasets, this can be of great advantage as it substantially reduces computational resources and time. Additionally, when new data becomes available, or the same data should be sorted with respect to different features, conventional neural networks require retraining, consuming significant computational memory and time. In contrast, in reservoir computing schemes, known reservoir results can be reused as indicated in Fig.~\ref{fig:RCscheme}; only potentially new input needs to be fed into the reservoir. The reservoir results of all data can then be used to train the computationally inexpensive linear regression or classifier model. Once the training process is completed, the latency of the model in predicting the output depends on the processing speed of the reservoir, making it suitable for exploiting the ultra-fast dynamics of nanoscale systems.

To function as a reservoir, a physical system must possess four key attributes~\cite{Jaeger2001b, Dambre2012, Inubushi2017, Love2023, Lee2023a}:
\begin{itemize}
\itemsep0em 
    \item non-linearity,
    \item complexity,
    \item short-term or fading memory, and
    \item reproducibility.
\end{itemize}
The non-linearity captures the degree to which the input is non-linearly transformed after passing through the reservoir, and the complexity property indicates how effectively the input is transformed into a higher-dimensional space. High dimensionality, in this context, means that the reservoir must have a significantly higher number of degrees of freedom compared to the input. Furthermore, because the tasks that reservoir computing commonly addresses are time-dependent, the reservoir must respond to the temporal history of the input signal. This is reflected in its short-term memory property, which prioritizes recent events while retaining past input history. Another crucial prerequisite, particularly in the context of physical reservoirs, is the need for reproducibility; the reservoir must consistently produce the same response when subjected to identical input conditions~\cite{Love2023, Cucchi2022}. 

Although reservoir computing can be implemented using transistor-based hardware, these traditional approaches are far from ideal because essential properties, like short-term memory, must be artificially imposed~\cite{Kudithipudi2016, Yi2016}. Moreover, the hardware system has to be specifically designed to emulate a randomly connected network, so that it can act as the reservoir. In contrast, many physical systems directly provide these properties, where the complexity and dynamical behavior can be naturally leveraged for computation~\cite{Tanaka2019, Nakajima2020, Christensen2022}. Firstly, for most physical systems, there are many internal degrees of freedom that dictate the dynamics, and consequently, the system does not need to be specifically configured. Secondly, any kind of damping, inertia, or dissipation ensures that even if the input driving the system is turned off, it takes some time for the system to return to its ground or metastable state. This plays the role of short-term memory or fading memory. Moreover, because these criteria do not specify the size of a reservoir, we can employ physical systems with nanometric or even atomic-level dimensions. Such use of nanoscale physical systems provides us with vast opportunities, potentially revolutionizing the scalability and efficiency of \textit{in-materio} computing.

In principle, many materials qualify as reservoirs \cite{Tanaka2019}. The challenge remains to identify the optimal system based on the specific application requirements. Primarily, reservoirs need to be practical, resilient, easy and cheap to fabricate, scalable, and energy efficient. Furthermore, they must operate within the time scale dictated by the problem at hand. The underlying physics of a reservoir gives its complexity, non-linearity, and fading memory time scales. Thus, a universal physical reservoir capable of handling all types of computation is not feasible, and this underscores the need for a variety or a combination of functional material systems and continuous research efforts to meet the evolving demands of information and communication technologies.

\section{Magnetic systems for reservoir computing}

Magnetic systems are suited as reservoirs due to their intrinsically complex and non-linear dynamical responses to various drives, including magnetic fields, electric currents, and light~\cite{Vedmedenko2020, Lee2023a, Grollier2020}. Fading memory naturally occurs because of the Gilbert damping and other dissipative sources in the magnetization dynamics. 
Furthermore, magnetic systems are CMOS compatible and therefore, magnetic reservoirs have the potential to be easily integrated into existing electronic devices.
Magnetic tunnel junctions, widely adopted in industry for their switch-like resistive property, are a prime example of this~\cite{Furuta2018, Finocchio2021}. However, magnetic tunnel junctions, as well as spin–vortex nano oscillators~\cite{Torrejon2017, Markovic2019}, do not provide the high complexity needed for reservoir computing. To use them as reservoirs, the complexity has to be artificially enhanced, for example, by using time multiplexing techniques~\cite{Cucchi2022, Van2017, Rohm2018}. Magnetic reservoirs which naturally provide a higher complexity include dipole-coupled nanomagnets~\cite{Nomura2019, Gartside2022}, spin wave-based reservoirs~\cite{Nomura2019}, magnetic metamaterials~\cite{Vidamour2023}, single skyrmion~\cite{Raab2022}, skyrmion fabrics~\cite{Prychynenko2018, Bourianoff2018, Pinna2018, Sun2023, LeeMK2022, LeeMK2023}, as well as other non-trivial spin textures~\cite{Bechler2023, Lee2023a, Lee2023b}. 

\begin{figure}[tb]
\includegraphics[width=1\columnwidth]{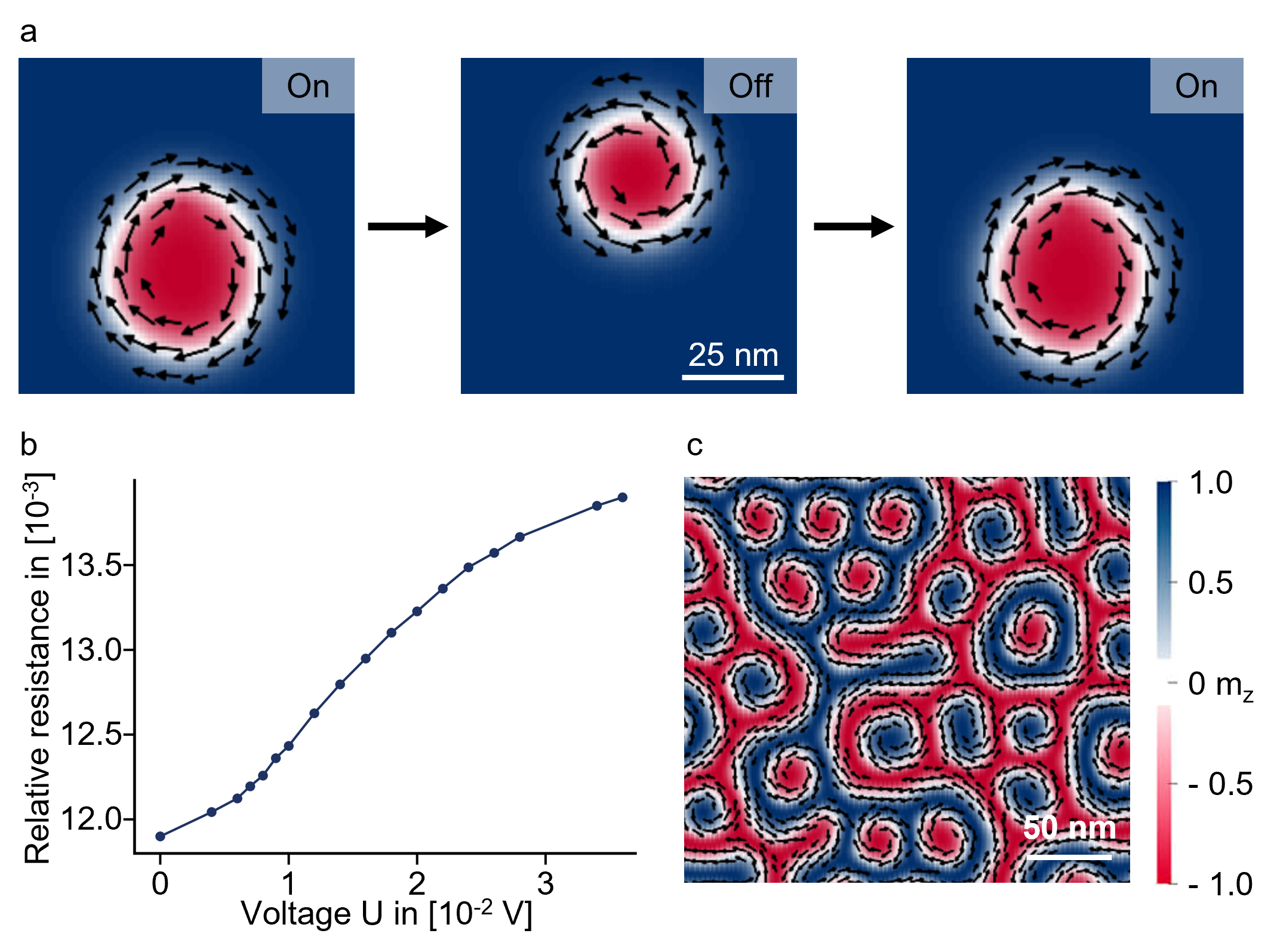}
\caption{Magnetic textures for reservoir computing obtained by micromagnetic simulations. Out-of-plane component ($m_z$) is shown by color and in-plane components by arrows. 
a) 
Single pinned skyrmion in a two-contact device (contacts are $\sim100$~nm away from the skyrmions and therefore not shown); turning the voltage off (on), the skyrmion relaxes by contracting (expands). 
b) Relative resistance changes as a function of applied voltage strength.
c) Skyrmion fabrics.
a) and b) are published in Ref.~\cite{Prychynenko2018} and c) is a modified versions from plots of Ref.~\cite{Bourianoff2018}.}
\label{fig:magnetic_system}
\end{figure}

One approach to reservoir computing using complex magnetic textures involves exciting them with an applied electrical voltage. In Figure~\ref{fig:magnetic_system}, the required properties of a reservoir are demonstrated using a skyrmion-based system~\cite{Prychynenko2018}. Applying a voltage with two contacts to a pinned skyrmion causes it to change its shape. This, in turn, causes a change in the electrical resistance owing to the anisotropic magnetoresistance (AMR) effect. The non-linear deformation (ON state) of the magnetic texture is caused by an interplay of the pinning strength and the spin-transfer torque. As shown in Fig.~\ref{fig:magnetic_system}(a), upon removing the positive voltage, the skyrmion responds by contracting to its metastable state (OFF state). Upon re-applying the voltage it undergoes the same deformation and expands in size to its previous state (ON state). This exhibits the reproducibility and fading memory properties of this system. Figure~\ref{fig:magnetic_system}(b) displays the non-linear response of the relative resistance to the applied voltage. These effects, and especially the complexity of the system, can be increased by using multiple skyrmions~\cite{Bourianoff2018, Pinna2018}, see Fig.~\ref{fig:magnetic_system}(c). 
The functionality and performance of reservoirs can be improved by, for example, using multiple contacts to supply voltage~\cite{Msiska2023} or by measuring both the spatial and the temporal variations of the observed quantity (in this case, the resistance)~\cite{Pinna2018}. 
Furthermore, extending the dimensionality of the magnetic system to the third spatial dimension promises higher information density and enables the use of 3D magnetic textures, such as (anti-)hedgehogs, chiral bobbers, extended domain walls, hopfions, and more.
A major advantage of using (antiferro-)magnetic systems is that their (ultra-)fast intrinsic time scales promote real-time computation. Given their versatility and the depth of understanding of the underlying physics, magnetic systems emerge as a promising platform for advancing physical reservoir computing research.

\section{ferroelectric systems for reservoir computing}

Going beyond the magnetic systems discussed so far, ferroelectrics have been investigated as promising candidate materials for reservoir computing and first device concepts have been proposed~\cite{Chen2023}. Ferroelectrics exhibit spontaneous long-range order of electric dipoles, which leads to an electric polarization. Analogous to the spins in magnets, the electric dipoles can readily be controlled by external stimuli, such as electric fields~\cite{Tagantsev2010}, light~\cite{Yang2018}, and stress~\cite{Lu2012}, giving rise to the non-linear and complex responses required for reservoir systems. Short-term memory arises from the relaxation of electric dipole arrangements after driving them away from equilibrium, e.g., by volatile ferroelectric switching. The idea of utilizing electric dipoles instead of magnetic spins is well-established and different concepts for reservoir computing have been proposed using, e.g., ferroelectric tunnel junctions (FTJs)~\cite{Kim2023}, ferroelectric field-effect transistors (FeFETs)~\cite{Tang2023, Toprasertpong2022}, and ferroelectric diodes~\cite{Chen2023}. An important leap ahead in this context was the discovery of ferroelectrics compatible with CMOS technology, enabling the design of densely integrated hardware \cite{Falcone2022}.

\begin{figure}[tb]
\includegraphics[width=0.9\columnwidth]{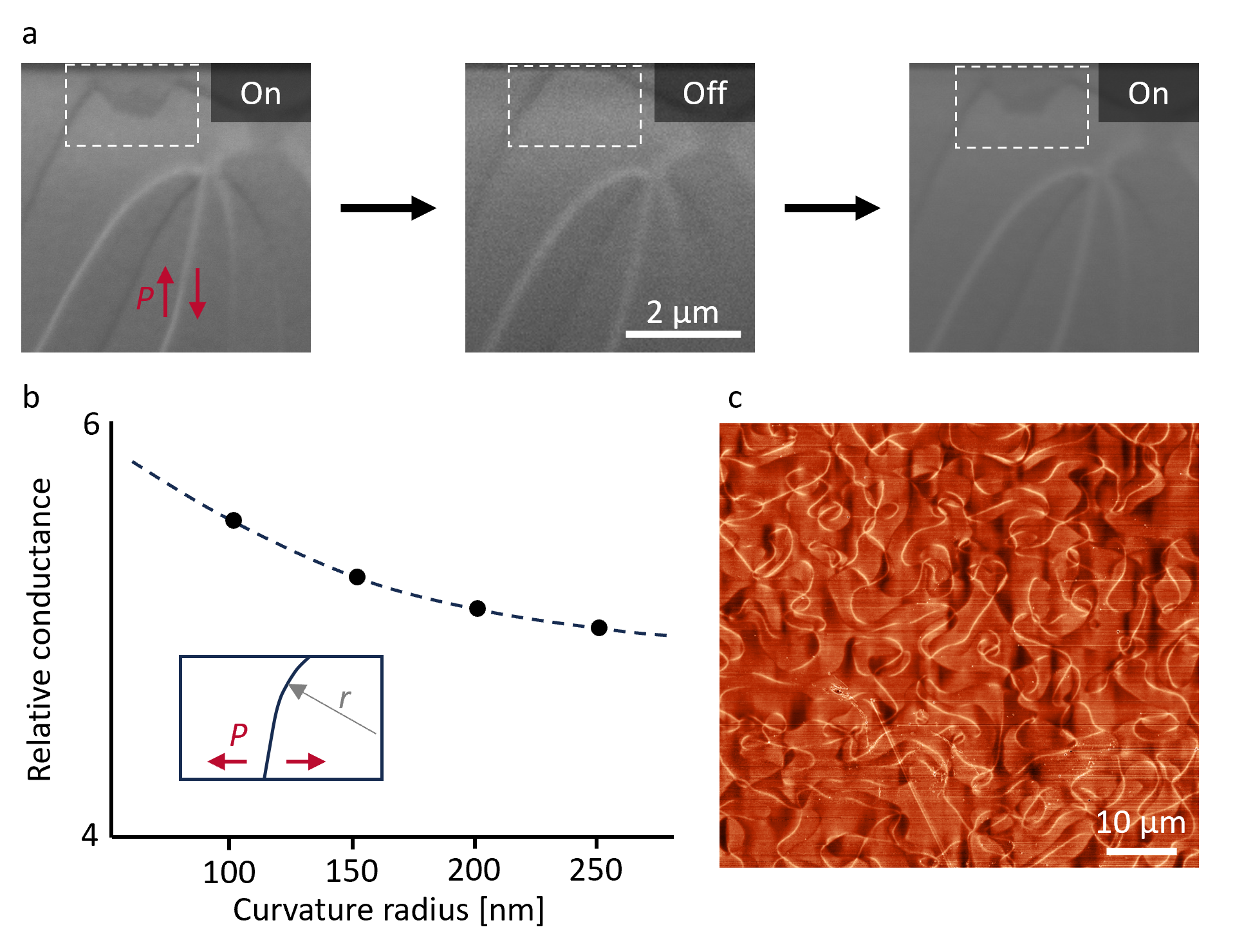}
\caption{Key features of ferroelectric domain walls for reservoir computing. a) Images series gained on ErMnO$_3$ by scanning electron microscopy adapted from~\cite{Roede2021}, showing that a local domain wall deformation within the region marked by the white dashed line can reversibly be induced by application of an electrical bias (on). As the applied electrical bias is removed, the domain wall relaxes back to its initial state (off). Red arrows indicate the polarization direction, $P$, in the ferroelectric domains. b) Sketch showing the non-linear relation between the relative conductance (with respect to the domains) of a concave tail-to-tail domain wall and the curvature radius $r$ as a function of the curvature radius, illustrated based on data published in~\cite{Roede2022}. c) Scanning probe microscopy image showing the extended network of domain walls that naturally forms in ErMnO$_3$ (courtesy of J. Schaab; EFM-2$\omega$ image, see~\cite{Schaab2015} for details)
}
\label{fig:ferroelectrics}
\end{figure}

Completely new and as-yet-unexplored opportunities for reservoir computing arise from topological defects in ferroelectrics~\cite{Nataf2020}, such as electric skyrmions and domain walls. Figure~\ref{fig:ferroelectrics} summarizes the key physical properties of ferroelectric domain walls, showcasing how the walls can be utilized for reservoir computing. By application of an electric voltage, ferroelectric domain walls can be moved and bent, which locally changes their charge state and, hence, gives rise to non-linear changes in conductance~\cite{Eliseev2011, Meier2012}. In Figure~\ref{fig:ferroelectrics}(a), reversible domain wall displacement is shown for the model system erbium manganite (ErMnO$_3$). Dark and bright lines correspond to positively (head-to-head) and negatively (tail-to-tail) charged domain walls, respectively, imaged by scanning electron microscopy. Positive charging temporarily deforms one of the domain walls in the region marked by the white dashed line (ON state), which relaxes back to its initial state (OFF state) as the applied electrical bias is removed~\cite{Roede2021}. This behavior gives the reproducibility and fading memory, analogous to the current-driven skyrmion deformation in Fig.~\ref{fig:magnetic_system}(a). Partial switching as seen in Fig.~\ref{fig:ferroelectrics}(a) can lead to pronounced temporary changes in conductance~\cite{Jiang2018}, as the domain wall charge state and conduction change together with the domain wall orientation~\cite{Meier2012}. This relation is displayed in Fig.~\ref{fig:ferroelectrics}(b) for the case of a conducting tail-to-tail wall in ErMnO$_3$, showing how the relative conductance changes as a function of the domain wall curvature. The latter can be controlled, e.g., by the application of an electric voltage to induce volatile switching as illustrated in Fig.~\ref{fig:ferroelectrics}(a), giving rise to the non-linear response required for reservoir computing. Reservoirs may be realized based on individual ferroelectric domain walls or domain wall networks as presented in Fig.~\ref{fig:ferroelectrics}(c). Here, the choice of the system that serves as the physical reservoir is crucial as the material needs to relax back to the initial domain structure to ensure reproducibility. In ErMnO$_3$, the latter is facilitated by topologically protected six-fold meeting points of domain walls, which prohibit complete poling and promote the relaxation back to the original domain structure~\cite{Choi2010, Han2013, Roede2021}. Just like for reservoirs based on magnetic skyrmions, the number of electrodes may be increased to induce domain wall deformations in multiple positions and both spatial and temporal changes in resistance can be recorded, enhancing the reservoir's performance.

Compared to the relaxation dynamics of magnetic skyrmions (Fig.~\ref{fig:magnetic_system}(a)), the transition between ON and OFF states in ferroelectric domain-wall-based reservoirs is expected to be much slower (the velocity of walls is limited by the speed of sound, with typical values in the order of 10 ms$^{-1}$~\cite{Grigoriev2006}), but their physical properties yield different advantages. For example, the walls in ErMnO$_3$ have a width of about 7~Å~\cite{Holtz2017}, allowing ultra-dense packing, and they naturally form a three-dimensional network that can readily be leveraged to increase the system's complexity and expand into third spatial dimension~\cite{Roede2022}. Furthermore, as domain wall displacements are voltage-controlled, there is no need for electrical currents, facilitating low power consumption and suppressing unwanted energy dissipation from Joule heating. With the discovery of new topological defects in ferroelectrics, including electric skyrmions~\cite{Das2019}, merons~\cite{Wang2020}, and hopfions~\cite{Luk2020}, application opportunities for ferroelectrics in reservoir computing continuously expand. The latter is propelled by close analogies between topological textures in magnetic and electric systems, which make device concepts transferable and enable a transition from spin to charge.

\begin{figure*}[t]
\includegraphics[width=0.8\linewidth]{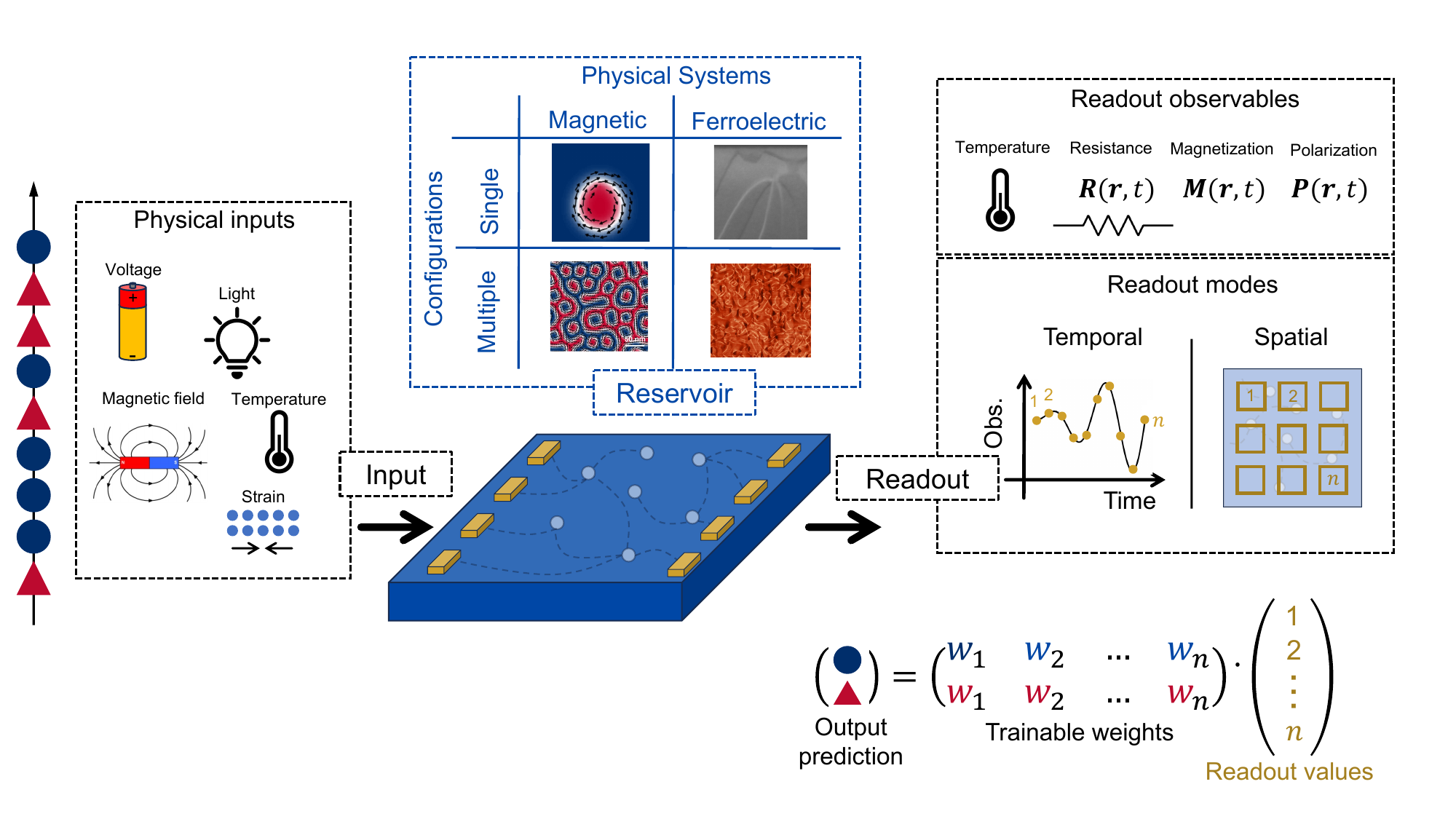}
\caption{Sketch of different physical operational / readout modes of reservoir computing.
a) temporal tracing of experimental signals; b) spatial measurement of the correlated reservoir's state.
Measured values serve as an input array to train a linear classifier sketched in c).}
\label{fig:operational_modes}
\end{figure*}

\section{Discussion and Outlook}

There is immense potential for physical reservoir computing in magnetic and ferroelectric systems, with its implementation across a vast array of modalities, as depicted in Fig.~\ref{fig:operational_modes}.
The physical system itself can consist of a single element (a solitary skyrmion or a single ferroelectric domain wall) or multiple co-existing textures (skyrmion fabrics or domain wall networks). 

Importantly, the respective system can be controlled by a range of physical inputs, such as light, voltage, magnetic field, temperature, and strain, giving the possibility to utilize it as a physical reservoir in different settings, ranging from nanoelectronics to optics. The external drive excites the system, which in turn can be read out using many different observables, such as resistance, temperature, susceptibility, magnetization, and polarization.

Apart from the different physical variables, different readout modes can be used. For example, the response of the reservoir can be read out by monitoring observables over time.
Alternatively, and even additionally, the readout can be done by spatially sampling the magnetic or electric systems.

The respective readout values are then used to train the weights of the final linear layer that yields the output predictions.
While in the current physical reservoir computers the training of weights in the final layer is carried out on an external device, there is still a need for research into how this step can also be carried out directly in the material. Another aspect that requires further research is the evaluation of a fair and reliable comparison of the performance of different reservoirs, including the careful separation of the influence of pre-processing, as well as the effects that arise when artificially increasing the complexity of a system~\cite{Torrejon2017, Msiska2023, Love2023}.
 
To summarize, magnetic and ferroelectric systems are very versatile and ideally suited for reservoir computing. The utilization of magnetic and electric topological defects (e.g., skyrmions and domain walls) gives a new dimension to the field of reservoir computing, facilitating the down-scaling of devices and enhanced complexity for improved performance. Furthermore, the very different natural time scales associated with spin and charge degrees of freedom in solid-state systems make them suitable for different applications. 

Interestingly, due to their complex structures which also extend non-trivially into the third dimension, topological defects in both magnetic and ferroelectric materials will make it possible to extend material-based reservoir computing to three dimensions and thus enable very compact computing units.

In addition, due to the large tunability of the systems realizing topological defects at different time and length scales, magnetic and ferroelectric structures naturally provide a wide range of applications. Given the co-existence of topological magnetic and electric defects in various systems, such as multiferroics~\cite{Fiebig2016} or superconducting-magnetic bilayer systems~\cite{Nothhelfer2022}, the complexity and range of possible physical inputs and readout observables can readily be further enhanced to broaden the range of applications. Moreover, connecting different physical reservoirs in a network structure allows for control and programmability of the reservoir~\cite{Stenning2022}. 

At this stage, however, it is fair to say that we have only come to the verge of understanding the unique opportunities for reservoir computing that arise from topological defects in magnetic and electric materials. It is an exciting time where new discoveries on the materials side almost go hand in hand with the development of innovative device concepts, boosted by the increasing interest of technology developers. The latter is reflected by commercially available neural network processors for edge computing and Internet of Things applications, suggesting that it might just be a question of time until the first physical reservoirs based on topological defects enter the market. The stunning progress within the field is facilitated by strong synergies, unifying researchers from physics, mathematics, materials science, and engineering. Importantly, research activities are still on the rise, with great potential to establish innovative approaches for the recognition and prediction of spatio-temporal events and future \textit{in-materio} computing in general.

\begin{acknowledgments}
We are grateful to Robin Msiska for numerous discussions.
K.E.S.\ acknowledges funding from the Emergent AI Center funded by the Carl Zeiss Foundation and the German Research Foundation (DFG) through the Emmy Noether grant under project No.~320163632. D.M.\ acknowledges funding from the European Research Council (ERC) under the European Union’s Horizon 2020 Research and Innovation Program (Grant Agreement No.~863691) and thanks NTNU for support through the Onsager Fellowship Program and the Outstanding Academic Fellow Program. The work was partly supported by the Research Council of Norway through its Centers of Excellence funding scheme, project number 262633, “QuSpin”.
\end{acknowledgments}

\bibliography{references1}

\begin{thebibliography}{78}%
\makeatletter
\providecommand \@ifxundefined [1]{%
 \@ifx{#1\undefined}
}%
\providecommand \@ifnum [1]{%
 \ifnum #1\expandafter \@firstoftwo
 \else \expandafter \@secondoftwo
 \fi
}%
\providecommand \@ifx [1]{%
 \ifx #1\expandafter \@firstoftwo
 \else \expandafter \@secondoftwo
 \fi
}%
\providecommand \natexlab [1]{#1}%
\providecommand \enquote  [1]{``#1''}%
\providecommand \bibnamefont  [1]{#1}%
\providecommand \bibfnamefont [1]{#1}%
\providecommand \citenamefont [1]{#1}%
\providecommand \href@noop [0]{\@secondoftwo}%
\providecommand \href [0]{\begingroup \@sanitize@url \@href}%
\providecommand \@href[1]{\@@startlink{#1}\@@href}%
\providecommand \@@href[1]{\endgroup#1\@@endlink}%
\providecommand \@sanitize@url [0]{\catcode `\\12\catcode `\$12\catcode
  `\&12\catcode `\#12\catcode `\^12\catcode `\_12\catcode `\%12\relax}%
\providecommand \@@startlink[1]{}%
\providecommand \@@endlink[0]{}%
\providecommand \url  [0]{\begingroup\@sanitize@url \@url }%
\providecommand \@url [1]{\endgroup\@href {#1}{\urlprefix }}%
\providecommand \urlprefix  [0]{URL }%
\providecommand \Eprint [0]{\href }%
\providecommand \doibase [0]{https://doi.org/}%
\providecommand \selectlanguage [0]{\@gobble}%
\providecommand \bibinfo  [0]{\@secondoftwo}%
\providecommand \bibfield  [0]{\@secondoftwo}%
\providecommand \translation [1]{[#1]}%
\providecommand \BibitemOpen [0]{}%
\providecommand \bibitemStop [0]{}%
\providecommand \bibitemNoStop [0]{.\EOS\space}%
\providecommand \EOS [0]{\spacefactor3000\relax}%
\providecommand \BibitemShut  [1]{\csname bibitem#1\endcsname}%
\let\auto@bib@innerbib\@empty
\bibitem [{\citenamefont {Ielmini}\ and\ \citenamefont
  {Wong}(2018)}]{Ielmini2018}%
  \BibitemOpen
  \bibfield  {author} {\bibinfo {author} {\bibfnamefont {D.}~\bibnamefont
  {Ielmini}}\ and\ \bibinfo {author} {\bibfnamefont {H.-S.~P.}\ \bibnamefont
  {Wong}},\ }\bibfield  {title} {\bibinfo {title} {{In-memory computing with
  resistive switching devices}},\ }\href
  {https://doi.org/https://doi.org/10.1038/s41928-018-0092-2} {\bibfield
  {journal} {\bibinfo  {journal} {Nature Electronics}\ }\textbf {\bibinfo
  {volume} {1}},\ \bibinfo {pages} {333} (\bibinfo {year} {2018})}\BibitemShut
  {NoStop}%
\bibitem [{\citenamefont {Indiveri}\ \emph {et~al.}(2011)\citenamefont
  {Indiveri}, \citenamefont {Linares-Barranco}, \citenamefont {Hamilton},
  \citenamefont {Schaik}, \citenamefont {Etienne-Cummings}, \citenamefont
  {Delbruck}, \citenamefont {Liu}, \citenamefont {Dudek}, \citenamefont
  {H{\"a}fliger}, \citenamefont {Renaud}, \citenamefont {Schemmel},
  \citenamefont {Cauwenberghs}, \citenamefont {Arthur}, \citenamefont {Hynna},
  \citenamefont {Folowosele}, \citenamefont {Saighi}, \citenamefont
  {Serrano-Gotarredona}, \citenamefont {Wijekoon}, \citenamefont {Wang},\ and\
  \citenamefont {Boahen}}]{Indiveri2011}%
  \BibitemOpen
  \bibfield  {author} {\bibinfo {author} {\bibfnamefont {G.}~\bibnamefont
  {Indiveri}}, \bibinfo {author} {\bibfnamefont {B.}~\bibnamefont
  {Linares-Barranco}}, \bibinfo {author} {\bibfnamefont {T.~J.}\ \bibnamefont
  {Hamilton}}, \bibinfo {author} {\bibfnamefont {A.~v.}\ \bibnamefont
  {Schaik}}, \bibinfo {author} {\bibfnamefont {R.}~\bibnamefont
  {Etienne-Cummings}}, \bibinfo {author} {\bibfnamefont {T.}~\bibnamefont
  {Delbruck}}, \bibinfo {author} {\bibfnamefont {S.-C.}\ \bibnamefont {Liu}},
  \bibinfo {author} {\bibfnamefont {P.}~\bibnamefont {Dudek}}, \bibinfo
  {author} {\bibfnamefont {P.}~\bibnamefont {H{\"a}fliger}}, \bibinfo {author}
  {\bibfnamefont {S.}~\bibnamefont {Renaud}}, \bibinfo {author} {\bibfnamefont
  {J.}~\bibnamefont {Schemmel}}, \bibinfo {author} {\bibfnamefont
  {G.}~\bibnamefont {Cauwenberghs}}, \bibinfo {author} {\bibfnamefont
  {J.}~\bibnamefont {Arthur}}, \bibinfo {author} {\bibfnamefont
  {K.}~\bibnamefont {Hynna}}, \bibinfo {author} {\bibfnamefont
  {F.}~\bibnamefont {Folowosele}}, \bibinfo {author} {\bibfnamefont
  {S.}~\bibnamefont {Saighi}}, \bibinfo {author} {\bibfnamefont
  {T.}~\bibnamefont {Serrano-Gotarredona}}, \bibinfo {author} {\bibfnamefont
  {J.}~\bibnamefont {Wijekoon}}, \bibinfo {author} {\bibfnamefont
  {Y.}~\bibnamefont {Wang}},\ and\ \bibinfo {author} {\bibfnamefont
  {K.}~\bibnamefont {Boahen}},\ }\bibfield  {title} {\bibinfo {title}
  {{Neuromorphic silicon neuron circuits}},\ }\href
  {https://doi.org/10.3389/fnins.2011.00073} {\bibfield  {journal} {\bibinfo
  {journal} {Frontiers in Neuroscience}\ }\textbf {\bibinfo {volume} {5}},\
  \bibinfo {pages} {73} (\bibinfo {year} {2011})}\BibitemShut {NoStop}%
\bibitem [{\citenamefont {Finocchio}\ \emph {et~al.}(2021)\citenamefont
  {Finocchio}, \citenamefont {{Di Ventra}}, \citenamefont {Camsari},
  \citenamefont {Everschor-Sitte}, \citenamefont {{Khalili Amiri}},\ and\
  \citenamefont {Zeng}}]{Finocchio2021}%
  \BibitemOpen
  \bibfield  {author} {\bibinfo {author} {\bibfnamefont {G.}~\bibnamefont
  {Finocchio}}, \bibinfo {author} {\bibfnamefont {M.}~\bibnamefont {{Di
  Ventra}}}, \bibinfo {author} {\bibfnamefont {K.~Y.}\ \bibnamefont {Camsari}},
  \bibinfo {author} {\bibfnamefont {K.}~\bibnamefont {Everschor-Sitte}},
  \bibinfo {author} {\bibfnamefont {P.}~\bibnamefont {{Khalili Amiri}}},\ and\
  \bibinfo {author} {\bibfnamefont {Z.}~\bibnamefont {Zeng}},\ }\bibfield
  {title} {\bibinfo {title} {{The promise of spintronics for unconventional
  computing}},\ }\href {https://doi.org/10.1016/j.jmmm.2020.167506} {\bibfield
  {journal} {\bibinfo  {journal} {Journal of Magnetism and Magnetic Materials}\
  }\textbf {\bibinfo {volume} {521}},\ \bibinfo {pages} {167506} (\bibinfo
  {year} {2021})}\BibitemShut {NoStop}%
\bibitem [{\citenamefont {Finocchio}\ \emph {et~al.}(2023)\citenamefont
  {Finocchio}, \citenamefont {Bandyopadhyay}, \citenamefont {Lin},
  \citenamefont {Pan}, \citenamefont {Yang}, \citenamefont {Tomasello},
  \citenamefont {Panagopoulos}, \citenamefont {Carpentieri}, \citenamefont
  {Puliafito}, \citenamefont {{\AA}kerman}, \citenamefont {Trivedi},
  \citenamefont {Mukhopadhyay}, \citenamefont {Roy}, \citenamefont {Sangwan},
  \citenamefont {Hersam}, \citenamefont {Giordano}, \citenamefont {Yang},
  \citenamefont {Grollier}, \citenamefont {Camsari}, \citenamefont {Mcmahon},
  \citenamefont {Datta}, \citenamefont {Incorvia}, \citenamefont {Friedman},
  \citenamefont {Cotofana}, \citenamefont {Ciubotaru}, \citenamefont {Chumak},
  \citenamefont {Naeemi}, \citenamefont {Kaushik}, \citenamefont {Zhu},
  \citenamefont {Wang}, \citenamefont {Koiller}, \citenamefont {Aguilar},
  \citenamefont {Temporão}, \citenamefont {Makasheva}, \citenamefont
  {Tordi-Sanial}, \citenamefont {Hasler}, \citenamefont {Levy}, \citenamefont
  {Roychowdhury}, \citenamefont {Ganguly}, \citenamefont {Ghosh}, \citenamefont
  {Rodriquez}, \citenamefont {Sunada}, \citenamefont {Evershor-Sitte},
  \citenamefont {Lal}, \citenamefont {Jadhav}, \citenamefont {Ventra},
  \citenamefont {Pershin}, \citenamefont {Tatsumura},\ and\ \citenamefont
  {Goto}}]{Finocchio2023}%
  \BibitemOpen
  \bibfield  {author} {\bibinfo {author} {\bibfnamefont {G.}~\bibnamefont
  {Finocchio}}, \bibinfo {author} {\bibfnamefont {S.}~\bibnamefont
  {Bandyopadhyay}}, \bibinfo {author} {\bibfnamefont {P.}~\bibnamefont {Lin}},
  \bibinfo {author} {\bibfnamefont {G.}~\bibnamefont {Pan}}, \bibinfo {author}
  {\bibfnamefont {J.~J.}\ \bibnamefont {Yang}}, \bibinfo {author}
  {\bibfnamefont {R.}~\bibnamefont {Tomasello}}, \bibinfo {author}
  {\bibfnamefont {C.}~\bibnamefont {Panagopoulos}}, \bibinfo {author}
  {\bibfnamefont {M.}~\bibnamefont {Carpentieri}}, \bibinfo {author}
  {\bibfnamefont {V.}~\bibnamefont {Puliafito}}, \bibinfo {author}
  {\bibfnamefont {H.}~\bibnamefont {{\AA}kerman}, \bibfnamefont
  {Johan~Takesue}}, \bibinfo {author} {\bibfnamefont {A.~R.}\ \bibnamefont
  {Trivedi}}, \bibinfo {author} {\bibfnamefont {S.}~\bibnamefont
  {Mukhopadhyay}}, \bibinfo {author} {\bibfnamefont {K.}~\bibnamefont {Roy}},
  \bibinfo {author} {\bibfnamefont {V.~K.}\ \bibnamefont {Sangwan}}, \bibinfo
  {author} {\bibfnamefont {M.~C.}\ \bibnamefont {Hersam}}, \bibinfo {author}
  {\bibfnamefont {A.}~\bibnamefont {Giordano}}, \bibinfo {author}
  {\bibfnamefont {H.}~\bibnamefont {Yang}}, \bibinfo {author} {\bibfnamefont
  {J.}~\bibnamefont {Grollier}}, \bibinfo {author} {\bibfnamefont
  {K.}~\bibnamefont {Camsari}}, \bibinfo {author} {\bibfnamefont
  {P.}~\bibnamefont {Mcmahon}}, \bibinfo {author} {\bibfnamefont
  {S.}~\bibnamefont {Datta}}, \bibinfo {author} {\bibfnamefont {J.~A.}\
  \bibnamefont {Incorvia}}, \bibinfo {author} {\bibfnamefont {J.}~\bibnamefont
  {Friedman}}, \bibinfo {author} {\bibfnamefont {S.}~\bibnamefont {Cotofana}},
  \bibinfo {author} {\bibfnamefont {F.}~\bibnamefont {Ciubotaru}}, \bibinfo
  {author} {\bibfnamefont {A.}~\bibnamefont {Chumak}}, \bibinfo {author}
  {\bibfnamefont {A.~J.}\ \bibnamefont {Naeemi}}, \bibinfo {author}
  {\bibfnamefont {B.~K.}\ \bibnamefont {Kaushik}}, \bibinfo {author}
  {\bibfnamefont {Y.}~\bibnamefont {Zhu}}, \bibinfo {author} {\bibfnamefont
  {K.}~\bibnamefont {Wang}}, \bibinfo {author} {\bibfnamefont {B.}~\bibnamefont
  {Koiller}}, \bibinfo {author} {\bibfnamefont {G.}~\bibnamefont {Aguilar}},
  \bibinfo {author} {\bibfnamefont {G.}~\bibnamefont {Temporão}}, \bibinfo
  {author} {\bibfnamefont {K.}~\bibnamefont {Makasheva}}, \bibinfo {author}
  {\bibfnamefont {A.}~\bibnamefont {Tordi-Sanial}}, \bibinfo {author}
  {\bibfnamefont {J.}~\bibnamefont {Hasler}}, \bibinfo {author} {\bibfnamefont
  {W.}~\bibnamefont {Levy}}, \bibinfo {author} {\bibfnamefont {V.}~\bibnamefont
  {Roychowdhury}}, \bibinfo {author} {\bibfnamefont {S.}~\bibnamefont
  {Ganguly}}, \bibinfo {author} {\bibfnamefont {A.}~\bibnamefont {Ghosh}},
  \bibinfo {author} {\bibfnamefont {D.}~\bibnamefont {Rodriquez}}, \bibinfo
  {author} {\bibfnamefont {S.}~\bibnamefont {Sunada}}, \bibinfo {author}
  {\bibfnamefont {K.}~\bibnamefont {Evershor-Sitte}}, \bibinfo {author}
  {\bibfnamefont {A.}~\bibnamefont {Lal}}, \bibinfo {author} {\bibfnamefont
  {S.}~\bibnamefont {Jadhav}}, \bibinfo {author} {\bibfnamefont {M.~D.}\
  \bibnamefont {Ventra}}, \bibinfo {author} {\bibfnamefont {Y.}~\bibnamefont
  {Pershin}}, \bibinfo {author} {\bibfnamefont {K.}~\bibnamefont {Tatsumura}},\
  and\ \bibinfo {author} {\bibfnamefont {H.}~\bibnamefont {Goto}},\ }\bibfield
  {title} {\bibinfo {title} {{Roadmap for unconventional computing with
  nanotechnology}},\ }\href {https://doi.org/10.48550/arXiv.2301.06727}
  {\bibfield  {journal} {\bibinfo  {journal} {arXiv:2301.06727}\ } (\bibinfo
  {year} {2023})}\BibitemShut {NoStop}%
\bibitem [{\citenamefont {Lee}\ \emph {et~al.}(2023{\natexlab{a}})\citenamefont
  {Lee}, \citenamefont {Msiska}, \citenamefont {Brems}, \citenamefont {Kläui},
  \citenamefont {Kurebayashi},\ and\ \citenamefont
  {Everschor-Sitte}}]{Lee2023a}%
  \BibitemOpen
  \bibfield  {author} {\bibinfo {author} {\bibfnamefont {O.}~\bibnamefont
  {Lee}}, \bibinfo {author} {\bibfnamefont {R.}~\bibnamefont {Msiska}},
  \bibinfo {author} {\bibfnamefont {M.~A.}\ \bibnamefont {Brems}}, \bibinfo
  {author} {\bibfnamefont {M.}~\bibnamefont {Kläui}}, \bibinfo {author}
  {\bibfnamefont {H.}~\bibnamefont {Kurebayashi}},\ and\ \bibinfo {author}
  {\bibfnamefont {K.}~\bibnamefont {Everschor-Sitte}},\ }\bibfield  {title}
  {\bibinfo {title} {{Perspective on unconventional computing using magnetic
  skyrmions}},\ }\href {https://doi.org/10.1063/5.0148469} {\bibfield
  {journal} {\bibinfo  {journal} {Applied Physics Letters}\ }\textbf {\bibinfo
  {volume} {122}},\ \bibinfo {pages} {260501} (\bibinfo {year}
  {2023}{\natexlab{a}})}\BibitemShut {NoStop}%
\bibitem [{\citenamefont {Fert}\ \emph {et~al.}(2017)\citenamefont {Fert},
  \citenamefont {Reyren},\ and\ \citenamefont {Cros}}]{Fert2017}%
  \BibitemOpen
  \bibfield  {author} {\bibinfo {author} {\bibfnamefont {A.}~\bibnamefont
  {Fert}}, \bibinfo {author} {\bibfnamefont {N.}~\bibnamefont {Reyren}},\ and\
  \bibinfo {author} {\bibfnamefont {V.}~\bibnamefont {Cros}},\ }\bibfield
  {title} {\bibinfo {title} {Magnetic skyrmions: advances in physics and
  potential applications},\ }\href {https://doi.org/10.1038/natrevmats.2017.31}
  {\bibfield  {journal} {\bibinfo  {journal} {Nature Reviews Materials}\
  }\textbf {\bibinfo {volume} {2}},\ \bibinfo {pages} {1} (\bibinfo {year}
  {2017})}\BibitemShut {NoStop}%
\bibitem [{\citenamefont {Nagaosa}\ and\ \citenamefont
  {Tokura}(2013)}]{Nagaosa2013}%
  \BibitemOpen
  \bibfield  {author} {\bibinfo {author} {\bibfnamefont {N.}~\bibnamefont
  {Nagaosa}}\ and\ \bibinfo {author} {\bibfnamefont {Y.}~\bibnamefont
  {Tokura}},\ }\bibfield  {title} {\bibinfo {title} {Topological properties and
  dynamics of magnetic skyrmions},\ }\href {https://doi.org/nnano.2013.243}
  {\bibfield  {journal} {\bibinfo  {journal} {Nature Nanotechnology}\ }\textbf
  {\bibinfo {volume} {8}},\ \bibinfo {pages} {899} (\bibinfo {year}
  {2013})}\BibitemShut {NoStop}%
\bibitem [{\citenamefont {Back}\ \emph {et~al.}(2020)\citenamefont {Back},
  \citenamefont {Cros}, \citenamefont {Ebert}, \citenamefont {Everschor-Sitte},
  \citenamefont {Fert}, \citenamefont {Garst}, \citenamefont {Ma},
  \citenamefont {Mankovsky}, \citenamefont {Monchesky}, \citenamefont
  {Mostovoy}, \citenamefont {Nagaosa}, \citenamefont {Parkin}, \citenamefont
  {Pfleiderer}, \citenamefont {Reyren}, \citenamefont {Rosch}, \citenamefont
  {Taguchi}, \citenamefont {Tokura}, \citenamefont {von Bergmann},\ and\
  \citenamefont {Zang}}]{Back2020}%
  \BibitemOpen
  \bibfield  {author} {\bibinfo {author} {\bibfnamefont {C.}~\bibnamefont
  {Back}}, \bibinfo {author} {\bibfnamefont {V.}~\bibnamefont {Cros}}, \bibinfo
  {author} {\bibfnamefont {H.}~\bibnamefont {Ebert}}, \bibinfo {author}
  {\bibfnamefont {K.}~\bibnamefont {Everschor-Sitte}}, \bibinfo {author}
  {\bibfnamefont {A.}~\bibnamefont {Fert}}, \bibinfo {author} {\bibfnamefont
  {M.}~\bibnamefont {Garst}}, \bibinfo {author} {\bibfnamefont
  {T.}~\bibnamefont {Ma}}, \bibinfo {author} {\bibfnamefont {S.}~\bibnamefont
  {Mankovsky}}, \bibinfo {author} {\bibfnamefont {T.}~\bibnamefont
  {Monchesky}}, \bibinfo {author} {\bibfnamefont {M.}~\bibnamefont {Mostovoy}},
  \bibinfo {author} {\bibfnamefont {N.}~\bibnamefont {Nagaosa}}, \bibinfo
  {author} {\bibfnamefont {S.}~\bibnamefont {Parkin}}, \bibinfo {author}
  {\bibfnamefont {C.}~\bibnamefont {Pfleiderer}}, \bibinfo {author}
  {\bibfnamefont {N.}~\bibnamefont {Reyren}}, \bibinfo {author} {\bibfnamefont
  {A.}~\bibnamefont {Rosch}}, \bibinfo {author} {\bibfnamefont
  {Y.}~\bibnamefont {Taguchi}}, \bibinfo {author} {\bibfnamefont
  {Y.}~\bibnamefont {Tokura}}, \bibinfo {author} {\bibfnamefont
  {K.}~\bibnamefont {von Bergmann}},\ and\ \bibinfo {author} {\bibfnamefont
  {J.}~\bibnamefont {Zang}},\ }\bibfield  {title} {\bibinfo {title} {{The 2020
  skyrmionics roadmap}},\ }\href {https://doi.org/10.1088/1361-6463/ab8418}
  {\bibfield  {journal} {\bibinfo  {journal} {Journal of Physics D: Applied
  Physics}\ }\textbf {\bibinfo {volume} {53}},\ \bibinfo {pages} {363001}
  (\bibinfo {year} {2020})}\BibitemShut {NoStop}%
\bibitem [{\citenamefont {Vedmedenko}\ \emph {et~al.}(2020)\citenamefont
  {Vedmedenko}, \citenamefont {Kawakami}, \citenamefont {Sheka}, \citenamefont
  {Gambardella}, \citenamefont {Kirilyuk}, \citenamefont {Hirohata},
  \citenamefont {Binek}, \citenamefont {Chubykalo-Fesenko}, \citenamefont
  {Sanvito}, \citenamefont {Kirby}, \citenamefont {Grollier}, \citenamefont
  {Everschor-Sitte}, \citenamefont {Kampfrath}, \citenamefont {You},\ and\
  \citenamefont {Berger}}]{Vedmedenko2020}%
  \BibitemOpen
  \bibfield  {author} {\bibinfo {author} {\bibfnamefont {E.~Y.}\ \bibnamefont
  {Vedmedenko}}, \bibinfo {author} {\bibfnamefont {R.~K.}\ \bibnamefont
  {Kawakami}}, \bibinfo {author} {\bibfnamefont {D.~D.}\ \bibnamefont {Sheka}},
  \bibinfo {author} {\bibfnamefont {P.}~\bibnamefont {Gambardella}}, \bibinfo
  {author} {\bibfnamefont {A.}~\bibnamefont {Kirilyuk}}, \bibinfo {author}
  {\bibfnamefont {A.}~\bibnamefont {Hirohata}}, \bibinfo {author}
  {\bibfnamefont {C.}~\bibnamefont {Binek}}, \bibinfo {author} {\bibfnamefont
  {O.}~\bibnamefont {Chubykalo-Fesenko}}, \bibinfo {author} {\bibfnamefont
  {S.}~\bibnamefont {Sanvito}}, \bibinfo {author} {\bibfnamefont {B.~J.}\
  \bibnamefont {Kirby}}, \bibinfo {author} {\bibfnamefont {J.}~\bibnamefont
  {Grollier}}, \bibinfo {author} {\bibfnamefont {K.}~\bibnamefont
  {Everschor-Sitte}}, \bibinfo {author} {\bibfnamefont {T.}~\bibnamefont
  {Kampfrath}}, \bibinfo {author} {\bibfnamefont {C.~Y.}\ \bibnamefont {You}},\
  and\ \bibinfo {author} {\bibfnamefont {A.}~\bibnamefont {Berger}},\
  }\bibfield  {title} {\bibinfo {title} {The 2020 magnetism roadmap},\ }\href
  {https://doi.org/10.1088/1361-6463/ab9d98} {\bibfield  {journal} {\bibinfo
  {journal} {Journal of Physics D: Applied Physics}\ }\textbf {\bibinfo
  {volume} {53}},\ \bibinfo {pages} {453001} (\bibinfo {year}
  {2020})}\BibitemShut {NoStop}%
\bibitem [{\citenamefont {Tang}\ \emph {et~al.}(2021)\citenamefont {Tang},
  \citenamefont {Wu}, \citenamefont {Wang}, \citenamefont {Kong}, \citenamefont
  {Lv}, \citenamefont {Wei}, \citenamefont {Zang}, \citenamefont {Tian},\ and\
  \citenamefont {Du}}]{Tang2021}%
  \BibitemOpen
  \bibfield  {author} {\bibinfo {author} {\bibfnamefont {J.}~\bibnamefont
  {Tang}}, \bibinfo {author} {\bibfnamefont {Y.}~\bibnamefont {Wu}}, \bibinfo
  {author} {\bibfnamefont {W.}~\bibnamefont {Wang}}, \bibinfo {author}
  {\bibfnamefont {L.}~\bibnamefont {Kong}}, \bibinfo {author} {\bibfnamefont
  {B.}~\bibnamefont {Lv}}, \bibinfo {author} {\bibfnamefont {W.}~\bibnamefont
  {Wei}}, \bibinfo {author} {\bibfnamefont {J.}~\bibnamefont {Zang}}, \bibinfo
  {author} {\bibfnamefont {M.}~\bibnamefont {Tian}},\ and\ \bibinfo {author}
  {\bibfnamefont {H.}~\bibnamefont {Du}},\ }\bibfield  {title} {\bibinfo
  {title} {Magnetic skyrmion bundles and their current-driven dynamics},\
  }\href {https://doi.org/10.1038/s41565-021-00954-9} {\bibfield  {journal}
  {\bibinfo  {journal} {Nature Nanotechnology}\ }\textbf {\bibinfo {volume}
  {16}},\ \bibinfo {pages} {1086} (\bibinfo {year} {2021})}\BibitemShut
  {NoStop}%
\bibitem [{\citenamefont {Wang}\ \emph {et~al.}(2019)\citenamefont {Wang},
  \citenamefont {Qaiumzadeh},\ and\ \citenamefont {Brataas}}]{Wang2019}%
  \BibitemOpen
  \bibfield  {author} {\bibinfo {author} {\bibfnamefont {X.}~\bibnamefont
  {Wang}}, \bibinfo {author} {\bibfnamefont {A.}~\bibnamefont {Qaiumzadeh}},\
  and\ \bibinfo {author} {\bibfnamefont {A.}~\bibnamefont {Brataas}},\
  }\bibfield  {title} {\bibinfo {title} {Current-driven dynamics of magnetic
  hopfions},\ }\href {https://doi.org/10.1103/PhysRevLett.123.147203}
  {\bibfield  {journal} {\bibinfo  {journal} {Physical Review Letters}\
  }\textbf {\bibinfo {volume} {123}},\ \bibinfo {pages} {147203} (\bibinfo
  {year} {2019})}\BibitemShut {NoStop}%
\bibitem [{\citenamefont {Kent}\ \emph {et~al.}(2021)\citenamefont {Kent},
  \citenamefont {Reynolds}, \citenamefont {Raftrey}, \citenamefont {Campbell},
  \citenamefont {Virasawmy}, \citenamefont {Dhuey}, \citenamefont {Chopdekar},
  \citenamefont {Hierro-Rodriguez}, \citenamefont {Sorrentino}, \citenamefont
  {Pereiro}, \citenamefont {Ferrer}, \citenamefont {Hellman}, \citenamefont
  {Sutcliffe},\ and\ \citenamefont {Fischer}}]{Kent2021}%
  \BibitemOpen
  \bibfield  {author} {\bibinfo {author} {\bibfnamefont {N.}~\bibnamefont
  {Kent}}, \bibinfo {author} {\bibfnamefont {N.}~\bibnamefont {Reynolds}},
  \bibinfo {author} {\bibfnamefont {D.}~\bibnamefont {Raftrey}}, \bibinfo
  {author} {\bibfnamefont {I.~T.~G.}\ \bibnamefont {Campbell}}, \bibinfo
  {author} {\bibfnamefont {S.}~\bibnamefont {Virasawmy}}, \bibinfo {author}
  {\bibfnamefont {S.}~\bibnamefont {Dhuey}}, \bibinfo {author} {\bibfnamefont
  {R.~V.}\ \bibnamefont {Chopdekar}}, \bibinfo {author} {\bibfnamefont
  {A.}~\bibnamefont {Hierro-Rodriguez}}, \bibinfo {author} {\bibfnamefont
  {A.}~\bibnamefont {Sorrentino}}, \bibinfo {author} {\bibfnamefont
  {E.}~\bibnamefont {Pereiro}}, \bibinfo {author} {\bibfnamefont
  {S.}~\bibnamefont {Ferrer}}, \bibinfo {author} {\bibfnamefont
  {F.}~\bibnamefont {Hellman}}, \bibinfo {author} {\bibfnamefont
  {P.}~\bibnamefont {Sutcliffe}},\ and\ \bibinfo {author} {\bibfnamefont
  {P.}~\bibnamefont {Fischer}},\ }\bibfield  {title} {\bibinfo {title}
  {{Creation and observation of Hopfions in magnetic multilayer systems}},\
  }\href {https://doi.org/10.1038/s41467-021-21846-5} {\bibfield  {journal}
  {\bibinfo  {journal} {Nature Communications}\ }\textbf {\bibinfo {volume}
  {12}},\ \bibinfo {pages} {1562} (\bibinfo {year} {2021})}\BibitemShut
  {NoStop}%
\bibitem [{\citenamefont {Azhar}\ \emph {et~al.}(2022)\citenamefont {Azhar},
  \citenamefont {Kravchuk},\ and\ \citenamefont {Garst}}]{Azhar2022}%
  \BibitemOpen
  \bibfield  {author} {\bibinfo {author} {\bibfnamefont {M.}~\bibnamefont
  {Azhar}}, \bibinfo {author} {\bibfnamefont {V.~P.}\ \bibnamefont
  {Kravchuk}},\ and\ \bibinfo {author} {\bibfnamefont {M.}~\bibnamefont
  {Garst}},\ }\bibfield  {title} {\bibinfo {title} {Screw dislocations in
  chiral magnets},\ }\href {https://doi.org/10.1103/PhysRevLett.128.157204}
  {\bibfield  {journal} {\bibinfo  {journal} {Physical Review Letters}\
  }\textbf {\bibinfo {volume} {128}},\ \bibinfo {pages} {157204} (\bibinfo
  {year} {2022})}\BibitemShut {NoStop}%
\bibitem [{\citenamefont {Stepanova}\ \emph {et~al.}(2021)\citenamefont
  {Stepanova}, \citenamefont {Masell}, \citenamefont {Lysne}, \citenamefont
  {Schoenherr}, \citenamefont {Köhler}, \citenamefont {Paulsen}, \citenamefont
  {Qaiumzadeh}, \citenamefont {Kanazawa}, \citenamefont {Rosch}, \citenamefont
  {Tokura}, \citenamefont {Brataas}, \citenamefont {Garst},\ and\ \citenamefont
  {Meier}}]{Stepanova2021}%
  \BibitemOpen
  \bibfield  {author} {\bibinfo {author} {\bibfnamefont {M.}~\bibnamefont
  {Stepanova}}, \bibinfo {author} {\bibfnamefont {J.}~\bibnamefont {Masell}},
  \bibinfo {author} {\bibfnamefont {E.}~\bibnamefont {Lysne}}, \bibinfo
  {author} {\bibfnamefont {P.}~\bibnamefont {Schoenherr}}, \bibinfo {author}
  {\bibfnamefont {L.}~\bibnamefont {Köhler}}, \bibinfo {author} {\bibfnamefont
  {M.}~\bibnamefont {Paulsen}}, \bibinfo {author} {\bibfnamefont
  {A.}~\bibnamefont {Qaiumzadeh}}, \bibinfo {author} {\bibfnamefont
  {N.}~\bibnamefont {Kanazawa}}, \bibinfo {author} {\bibfnamefont
  {A.}~\bibnamefont {Rosch}}, \bibinfo {author} {\bibfnamefont
  {Y.}~\bibnamefont {Tokura}}, \bibinfo {author} {\bibfnamefont
  {A.}~\bibnamefont {Brataas}}, \bibinfo {author} {\bibfnamefont
  {M.}~\bibnamefont {Garst}},\ and\ \bibinfo {author} {\bibfnamefont
  {D.}~\bibnamefont {Meier}},\ }\bibfield  {title} {\bibinfo {title}
  {{Detection of Topological Spin Textures via Nonlinear Magnetic Responses}},\
  }\href {https://doi.org/10.1021/acs.nanolett.1c02723} {\bibfield  {journal}
  {\bibinfo  {journal} {Nano Letters}\ }\textbf {\bibinfo {volume} {22}},\
  \bibinfo {pages} {14} (\bibinfo {year} {2021})}\BibitemShut {NoStop}%
\bibitem [{\citenamefont {McConville}\ \emph {et~al.}(2020)\citenamefont
  {McConville}, \citenamefont {Lu}, \citenamefont {Wang}, \citenamefont {Tan},
  \citenamefont {Cochard}, \citenamefont {Conroy}, \citenamefont {Moore},
  \citenamefont {Harvey}, \citenamefont {Bangert}, \citenamefont {Chen},
  \citenamefont {Gruverman},\ and\ \citenamefont {Gregg}}]{Mcconville2020}%
  \BibitemOpen
  \bibfield  {author} {\bibinfo {author} {\bibfnamefont {J.~P.}\ \bibnamefont
  {McConville}}, \bibinfo {author} {\bibfnamefont {H.}~\bibnamefont {Lu}},
  \bibinfo {author} {\bibfnamefont {B.}~\bibnamefont {Wang}}, \bibinfo {author}
  {\bibfnamefont {Y.}~\bibnamefont {Tan}}, \bibinfo {author} {\bibfnamefont
  {C.}~\bibnamefont {Cochard}}, \bibinfo {author} {\bibfnamefont
  {M.}~\bibnamefont {Conroy}}, \bibinfo {author} {\bibfnamefont
  {K.}~\bibnamefont {Moore}}, \bibinfo {author} {\bibfnamefont
  {A.}~\bibnamefont {Harvey}}, \bibinfo {author} {\bibfnamefont
  {U.}~\bibnamefont {Bangert}}, \bibinfo {author} {\bibfnamefont {L.~Q.}\
  \bibnamefont {Chen}}, \bibinfo {author} {\bibfnamefont {A.}~\bibnamefont
  {Gruverman}},\ and\ \bibinfo {author} {\bibfnamefont {J.~M.}\ \bibnamefont
  {Gregg}},\ }\bibfield  {title} {\bibinfo {title} {{Ferroelectric domain wall
  memristor}},\ }\href {https://doi.org/10.1002/adfm.202000109} {\bibfield
  {journal} {\bibinfo  {journal} {Advanced Functional aterials}\ }\textbf
  {\bibinfo {volume} {30}},\ \bibinfo {pages} {2000109} (\bibinfo {year}
  {2020})}\BibitemShut {NoStop}%
\bibitem [{\citenamefont {Rieck}\ \emph {et~al.}(2023)\citenamefont {Rieck},
  \citenamefont {Cipollini}, \citenamefont {Salverda}, \citenamefont
  {Quinteros}, \citenamefont {Schomaker},\ and\ \citenamefont
  {Noheda}}]{Rieck2023}%
  \BibitemOpen
  \bibfield  {author} {\bibinfo {author} {\bibfnamefont {J.~L.}\ \bibnamefont
  {Rieck}}, \bibinfo {author} {\bibfnamefont {D.}~\bibnamefont {Cipollini}},
  \bibinfo {author} {\bibfnamefont {M.}~\bibnamefont {Salverda}}, \bibinfo
  {author} {\bibfnamefont {C.~P.}\ \bibnamefont {Quinteros}}, \bibinfo {author}
  {\bibfnamefont {L.~R.}\ \bibnamefont {Schomaker}},\ and\ \bibinfo {author}
  {\bibfnamefont {B.}~\bibnamefont {Noheda}},\ }\bibfield  {title} {\bibinfo
  {title} {Ferroelastic domain walls in {B}i{F}e{O$_3$} as memristive
  networks},\ }\href {https://doi.org/10.1002/aisy.202200292} {\bibfield
  {journal} {\bibinfo  {journal} {Advanced Intelligent Systems}\ }\textbf
  {\bibinfo {volume} {5}},\ \bibinfo {pages} {2200292} (\bibinfo {year}
  {2023})}\BibitemShut {NoStop}%
\bibitem [{\citenamefont {Meier}\ and\ \citenamefont
  {Selbach}(2022)}]{Meier2022}%
  \BibitemOpen
  \bibfield  {author} {\bibinfo {author} {\bibfnamefont {D.}~\bibnamefont
  {Meier}}\ and\ \bibinfo {author} {\bibfnamefont {S.~M.}\ \bibnamefont
  {Selbach}},\ }\bibfield  {title} {\bibinfo {title} {Ferroelectric domain
  walls for nanotechnology},\ }\href
  {https://doi.org/10.1038/s41578-021-00375-z} {\bibfield  {journal} {\bibinfo
  {journal} {Nature Reviews Materials}\ }\textbf {\bibinfo {volume} {7}},\
  \bibinfo {pages} {157} (\bibinfo {year} {2022})}\BibitemShut {NoStop}%
\bibitem [{\citenamefont {Sharma}\ \emph {et~al.}(2022)\citenamefont {Sharma},
  \citenamefont {Moise}, \citenamefont {Colombo},\ and\ \citenamefont
  {Seidel}}]{Sharma2022}%
  \BibitemOpen
  \bibfield  {author} {\bibinfo {author} {\bibfnamefont {P.}~\bibnamefont
  {Sharma}}, \bibinfo {author} {\bibfnamefont {T.~S.}\ \bibnamefont {Moise}},
  \bibinfo {author} {\bibfnamefont {L.}~\bibnamefont {Colombo}},\ and\ \bibinfo
  {author} {\bibfnamefont {J.}~\bibnamefont {Seidel}},\ }\bibfield  {title}
  {\bibinfo {title} {Roadmap for ferroelectric domain wall nanoelectronics},\
  }\href {https://doi.org/10.1002/adfm.202110263} {\bibfield  {journal}
  {\bibinfo  {journal} {Advanced Functional Materials}\ }\textbf {\bibinfo
  {volume} {32}},\ \bibinfo {pages} {2110263} (\bibinfo {year}
  {2022})}\BibitemShut {NoStop}%
\bibitem [{\citenamefont {Wang}\ \emph {et~al.}(2022)\citenamefont {Wang},
  \citenamefont {Wang}, \citenamefont {Zhang}, \citenamefont {Jiang},
  \citenamefont {Chen},\ and\ \citenamefont {Jiang}}]{Wang2022}%
  \BibitemOpen
  \bibfield  {author} {\bibinfo {author} {\bibfnamefont {C.}~\bibnamefont
  {Wang}}, \bibinfo {author} {\bibfnamefont {T.}~\bibnamefont {Wang}}, \bibinfo
  {author} {\bibfnamefont {W.}~\bibnamefont {Zhang}}, \bibinfo {author}
  {\bibfnamefont {J.}~\bibnamefont {Jiang}}, \bibinfo {author} {\bibfnamefont
  {L.}~\bibnamefont {Chen}},\ and\ \bibinfo {author} {\bibfnamefont
  {A.}~\bibnamefont {Jiang}},\ }\bibfield  {title} {\bibinfo {title} {Analog
  ferroelectric domain-wall memories and synaptic devices integrated with {S}i
  substrates},\ }\href {https://doi.org/10.1007/s12274-021-3899-5} {\bibfield
  {journal} {\bibinfo  {journal} {Nano Research}\ }\textbf {\bibinfo {volume}
  {15}},\ \bibinfo {pages} {3606} (\bibinfo {year} {2022})}\BibitemShut
  {NoStop}%
\bibitem [{\citenamefont {Schroeder}\ \emph {et~al.}(2013)\citenamefont
  {Schroeder}, \citenamefont {Mueller}, \citenamefont {Mueller}, \citenamefont
  {Yurchuk}, \citenamefont {Martin}, \citenamefont {Adelmann}, \citenamefont
  {Schloesser}, \citenamefont {van Bentum},\ and\ \citenamefont
  {Mikolajick}}]{Schroeder2013}%
  \BibitemOpen
  \bibfield  {author} {\bibinfo {author} {\bibfnamefont {U.}~\bibnamefont
  {Schroeder}}, \bibinfo {author} {\bibfnamefont {S.}~\bibnamefont {Mueller}},
  \bibinfo {author} {\bibfnamefont {J.}~\bibnamefont {Mueller}}, \bibinfo
  {author} {\bibfnamefont {E.}~\bibnamefont {Yurchuk}}, \bibinfo {author}
  {\bibfnamefont {D.}~\bibnamefont {Martin}}, \bibinfo {author} {\bibfnamefont
  {C.}~\bibnamefont {Adelmann}}, \bibinfo {author} {\bibfnamefont
  {T.}~\bibnamefont {Schloesser}}, \bibinfo {author} {\bibfnamefont
  {R.}~\bibnamefont {van Bentum}},\ and\ \bibinfo {author} {\bibfnamefont
  {T.}~\bibnamefont {Mikolajick}},\ }\bibfield  {title} {\bibinfo {title}
  {{Hafnium oxide based CMOS compatible ferroelectric materials}},\ }\href
  {https://doi.org/10.1149/2.010304jss} {\bibfield  {journal} {\bibinfo
  {journal} {ECS Journal of Solid State Science and Technology}\ }\textbf
  {\bibinfo {volume} {2}},\ \bibinfo {pages} {N69} (\bibinfo {year}
  {2013})}\BibitemShut {NoStop}%
\bibitem [{\citenamefont {Govinden}\ \emph {et~al.}(2023)\citenamefont
  {Govinden}, \citenamefont {Prokhorenko}, \citenamefont {Zhang}, \citenamefont
  {Rijal}, \citenamefont {Nahas}, \citenamefont {Bellaiche},\ and\
  \citenamefont {Valanoor}}]{Govinden2023}%
  \BibitemOpen
  \bibfield  {author} {\bibinfo {author} {\bibfnamefont {V.}~\bibnamefont
  {Govinden}}, \bibinfo {author} {\bibfnamefont {S.}~\bibnamefont
  {Prokhorenko}}, \bibinfo {author} {\bibfnamefont {Q.}~\bibnamefont {Zhang}},
  \bibinfo {author} {\bibfnamefont {S.}~\bibnamefont {Rijal}}, \bibinfo
  {author} {\bibfnamefont {Y.}~\bibnamefont {Nahas}}, \bibinfo {author}
  {\bibfnamefont {L.}~\bibnamefont {Bellaiche}},\ and\ \bibinfo {author}
  {\bibfnamefont {N.}~\bibnamefont {Valanoor}},\ }\bibfield  {title} {\bibinfo
  {title} {Spherical ferroelectric solitons},\ }\href
  {https://doi.org/10.1038/s41563-023-01527-y} {\bibfield  {journal} {\bibinfo
  {journal} {Nature Materials}\ }\textbf {\bibinfo {volume} {22}},\ \bibinfo
  {pages} {553–561} (\bibinfo {year} {2023})}\BibitemShut {NoStop}%
\bibitem [{\citenamefont {Nataf}\ \emph {et~al.}(2020)\citenamefont {Nataf},
  \citenamefont {Guennou}, \citenamefont {Gregg}, \citenamefont {Meier},
  \citenamefont {Hlinka}, \citenamefont {Salje},\ and\ \citenamefont
  {Kreisel}}]{Nataf2020}%
  \BibitemOpen
  \bibfield  {author} {\bibinfo {author} {\bibfnamefont {G.~F.}\ \bibnamefont
  {Nataf}}, \bibinfo {author} {\bibfnamefont {M.}~\bibnamefont {Guennou}},
  \bibinfo {author} {\bibfnamefont {J.~M.}\ \bibnamefont {Gregg}}, \bibinfo
  {author} {\bibfnamefont {D.}~\bibnamefont {Meier}}, \bibinfo {author}
  {\bibfnamefont {J.}~\bibnamefont {Hlinka}}, \bibinfo {author} {\bibfnamefont
  {E.~K.}\ \bibnamefont {Salje}},\ and\ \bibinfo {author} {\bibfnamefont
  {J.}~\bibnamefont {Kreisel}},\ }\bibfield  {title} {\bibinfo {title}
  {Domain-wall engineering and topological defects in ferroelectric and
  ferroelastic materials},\ }\href {https://doi.org/10.1038/s42254-020-0235-z}
  {\bibfield  {journal} {\bibinfo  {journal} {Nature Reviews Physics}\ }\textbf
  {\bibinfo {volume} {2}},\ \bibinfo {pages} {634–648} (\bibinfo {year}
  {2020})}\BibitemShut {NoStop}%
\bibitem [{\citenamefont {Das}\ \emph {et~al.}(2019)\citenamefont {Das},
  \citenamefont {Tang}, \citenamefont {Hong}, \citenamefont {Gon\c{c}alves},
  \citenamefont {McCarter}, \citenamefont {Klewe}, \citenamefont {Nguyen},
  \citenamefont {G\'{o}mez-Ortiz}, \citenamefont {Shafer}, \citenamefont
  {Arenholz}, \citenamefont {Stoica}, \citenamefont {Hsu}, \citenamefont
  {Wang}, \citenamefont {Ophus}, \citenamefont {Liu}, \citenamefont {Nelson},
  \citenamefont {Saremi}, \citenamefont {Prasad}, \citenamefont {Mei},
  \citenamefont {Schlom}, \citenamefont {\'{I}\~{n}iguez}, \citenamefont
  {Garc\'{i}a-Fern\'{a}ndez}, \citenamefont {Muller}, \citenamefont {Chen},
  \citenamefont {Junquera}, \citenamefont {Martin},\ and\ \citenamefont
  {Ramesh}}]{Das2019}%
  \BibitemOpen
  \bibfield  {author} {\bibinfo {author} {\bibfnamefont {S.}~\bibnamefont
  {Das}}, \bibinfo {author} {\bibfnamefont {Y.~L.}\ \bibnamefont {Tang}},
  \bibinfo {author} {\bibfnamefont {Z.}~\bibnamefont {Hong}}, \bibinfo {author}
  {\bibfnamefont {M.~A.~P.}\ \bibnamefont {Gon\c{c}alves}}, \bibinfo {author}
  {\bibfnamefont {M.~R.}\ \bibnamefont {McCarter}}, \bibinfo {author}
  {\bibfnamefont {C.}~\bibnamefont {Klewe}}, \bibinfo {author} {\bibfnamefont
  {K.~X.}\ \bibnamefont {Nguyen}}, \bibinfo {author} {\bibfnamefont
  {F.}~\bibnamefont {G\'{o}mez-Ortiz}}, \bibinfo {author} {\bibfnamefont
  {P.}~\bibnamefont {Shafer}}, \bibinfo {author} {\bibfnamefont
  {E.}~\bibnamefont {Arenholz}}, \bibinfo {author} {\bibfnamefont {V.~A.}\
  \bibnamefont {Stoica}}, \bibinfo {author} {\bibfnamefont {S.-L.}\
  \bibnamefont {Hsu}}, \bibinfo {author} {\bibfnamefont {B.}~\bibnamefont
  {Wang}}, \bibinfo {author} {\bibfnamefont {C.}~\bibnamefont {Ophus}},
  \bibinfo {author} {\bibfnamefont {J.~F.}\ \bibnamefont {Liu}}, \bibinfo
  {author} {\bibfnamefont {C.~T.}\ \bibnamefont {Nelson}}, \bibinfo {author}
  {\bibfnamefont {S.}~\bibnamefont {Saremi}}, \bibinfo {author} {\bibfnamefont
  {B.}~\bibnamefont {Prasad}}, \bibinfo {author} {\bibfnamefont {A.~B.}\
  \bibnamefont {Mei}}, \bibinfo {author} {\bibfnamefont {D.~G.}\ \bibnamefont
  {Schlom}}, \bibinfo {author} {\bibfnamefont {J.}~\bibnamefont
  {\'{I}\~{n}iguez}}, \bibinfo {author} {\bibfnamefont {P.}~\bibnamefont
  {Garc\'{i}a-Fern\'{a}ndez}}, \bibinfo {author} {\bibfnamefont {D.~A.}\
  \bibnamefont {Muller}}, \bibinfo {author} {\bibfnamefont {L.~Q.}\
  \bibnamefont {Chen}}, \bibinfo {author} {\bibfnamefont {J.}~\bibnamefont
  {Junquera}}, \bibinfo {author} {\bibfnamefont {L.~W.}\ \bibnamefont
  {Martin}},\ and\ \bibinfo {author} {\bibfnamefont {R.}~\bibnamefont
  {Ramesh}},\ }\bibfield  {title} {\bibinfo {title} {Observation of
  room-temperature polar skyrmions},\ }\href
  {https://doi.org/10.1038/s41586-019-1092-8} {\bibfield  {journal} {\bibinfo
  {journal} {Nature}\ }\textbf {\bibinfo {volume} {568}},\ \bibinfo {pages}
  {368} (\bibinfo {year} {2019})}\BibitemShut {NoStop}%
\bibitem [{Foc()}]{Focusissue}%
  \BibitemOpen
  \href
  {https://iopscience.iop.org/journal/2634-4386/page/focus-issue-on-topological-solitons-for-neuromorphic-systems}
  {\bibfield  {journal} {\bibinfo  {journal} {Focus issue on Topological
  Solitons for Neuromorphic Systems}\ }}\bibinfo {note} {Accessed:
  2023-05-10}\BibitemShut {NoStop}%
\bibitem [{\citenamefont {Luko{\v{s}}evi{\v{c}}ius}\ \emph
  {et~al.}(2012)\citenamefont {Luko{\v{s}}evi{\v{c}}ius}, \citenamefont
  {Jaeger},\ and\ \citenamefont {Schrauwen}}]{Lukosevicius2012}%
  \BibitemOpen
  \bibfield  {author} {\bibinfo {author} {\bibfnamefont {M.}~\bibnamefont
  {Luko{\v{s}}evi{\v{c}}ius}}, \bibinfo {author} {\bibfnamefont
  {H.}~\bibnamefont {Jaeger}},\ and\ \bibinfo {author} {\bibfnamefont
  {B.}~\bibnamefont {Schrauwen}},\ }\bibfield  {title} {\bibinfo {title}
  {{Reservoir Computing Trends}},\ }\href
  {https://doi.org/10.1007/s13218-012-0204-5} {\bibfield  {journal} {\bibinfo
  {journal} {KI - K{\"{u}}nstliche Intelligenz}\ }\textbf {\bibinfo {volume}
  {26}},\ \bibinfo {pages} {365} (\bibinfo {year} {2012})}\BibitemShut
  {NoStop}%
\bibitem [{\citenamefont {Jaeger}(2001{\natexlab{a}})}]{Jaeger2001}%
  \BibitemOpen
  \bibfield  {author} {\bibinfo {author} {\bibfnamefont {H.}~\bibnamefont
  {Jaeger}},\ }\bibfield  {title} {\bibinfo {title} {The “echo state”
  approach to analysing and training recurrent neural networks-with an erratum
  note},\ }\href
  {https://doi.org/https://www.ai.rug.nl/minds/uploads/EchoStatesTechRep.pdf}
  {\bibfield  {journal} {\bibinfo  {journal} {Bonn, Germany: German National
  Research Center for Information Technology GMD Technical Report}\ }\textbf
  {\bibinfo {volume} {148}},\ \bibinfo {pages} {13} (\bibinfo {year}
  {2001}{\natexlab{a}})}\BibitemShut {NoStop}%
\bibitem [{\citenamefont {Jaeger}(2001{\natexlab{b}})}]{Jaeger2001b}%
  \BibitemOpen
  \bibfield  {author} {\bibinfo {author} {\bibfnamefont {H.}~\bibnamefont
  {Jaeger}},\ }\bibfield  {title} {\bibinfo {title} {Short term memory in echo
  state networks},\ }\href
  {https://publica.fraunhofer.de/entities/publication/9dfaead1-4dc0-4e3c-b89b-596f50f671c1/details}
  {\bibfield  {journal} {\bibinfo  {journal} {GMD Forschungszentrum
  Informationstechnik}\ } (\bibinfo {year} {2001}{\natexlab{b}})}\BibitemShut
  {NoStop}%
\bibitem [{\citenamefont {Dambre}\ \emph {et~al.}(2012)\citenamefont {Dambre},
  \citenamefont {Verstraeten}, \citenamefont {Schrauwen},\ and\ \citenamefont
  {Massar}}]{Dambre2012}%
  \BibitemOpen
  \bibfield  {author} {\bibinfo {author} {\bibfnamefont {J.}~\bibnamefont
  {Dambre}}, \bibinfo {author} {\bibfnamefont {D.}~\bibnamefont {Verstraeten}},
  \bibinfo {author} {\bibfnamefont {B.}~\bibnamefont {Schrauwen}},\ and\
  \bibinfo {author} {\bibfnamefont {S.}~\bibnamefont {Massar}},\ }\bibfield
  {title} {\bibinfo {title} {Information processing capacity of dynamical
  systems},\ }\href {https://doi.org/https://doi.org/10.1038/srep00514}
  {\bibfield  {journal} {\bibinfo  {journal} {Scientific Reports}\ }\textbf
  {\bibinfo {volume} {2}},\ \bibinfo {pages} {514} (\bibinfo {year}
  {2012})}\BibitemShut {NoStop}%
\bibitem [{\citenamefont {Inubushi}\ and\ \citenamefont
  {Yoshimura}(2017)}]{Inubushi2017}%
  \BibitemOpen
  \bibfield  {author} {\bibinfo {author} {\bibfnamefont {M.}~\bibnamefont
  {Inubushi}}\ and\ \bibinfo {author} {\bibfnamefont {K.}~\bibnamefont
  {Yoshimura}},\ }\bibfield  {title} {\bibinfo {title} {Reservoir computing
  beyond memory-nonlinearity trade-off},\ }\href
  {https://doi.org/https://doi.org/10.1038/s41598-017-10257-6} {\bibfield
  {journal} {\bibinfo  {journal} {Scientific Reports}\ }\textbf {\bibinfo
  {volume} {7}},\ \bibinfo {pages} {10199} (\bibinfo {year}
  {2017})}\BibitemShut {NoStop}%
\bibitem [{\citenamefont {Love}\ \emph {et~al.}(2023)\citenamefont {Love},
  \citenamefont {Msiska}, \citenamefont {Mulkers}, \citenamefont {Bourianoff},
  \citenamefont {Leliaert},\ and\ \citenamefont {Everschor-Sitte}}]{Love2023}%
  \BibitemOpen
  \bibfield  {author} {\bibinfo {author} {\bibfnamefont {J.}~\bibnamefont
  {Love}}, \bibinfo {author} {\bibfnamefont {R.}~\bibnamefont {Msiska}},
  \bibinfo {author} {\bibfnamefont {J.}~\bibnamefont {Mulkers}}, \bibinfo
  {author} {\bibfnamefont {G.}~\bibnamefont {Bourianoff}}, \bibinfo {author}
  {\bibfnamefont {J.}~\bibnamefont {Leliaert}},\ and\ \bibinfo {author}
  {\bibfnamefont {K.}~\bibnamefont {Everschor-Sitte}},\ }\bibfield  {title}
  {\bibinfo {title} {Spatial analysis of physical reservoir computers},\ }\href
  {https://doi.org/https://doi.org/10.1103/PhysRevApplied.20.044057} {\bibfield
   {journal} {\bibinfo  {journal} {Physical Review Applied}\ }\textbf {\bibinfo
  {volume} {20}},\ \bibinfo {pages} {044057} (\bibinfo {year}
  {2023})}\BibitemShut {NoStop}%
\bibitem [{\citenamefont {Cucchi}\ \emph {et~al.}(2022)\citenamefont {Cucchi},
  \citenamefont {Abreu}, \citenamefont {Ciccone}, \citenamefont {Brunner},\
  and\ \citenamefont {Kleemann}}]{Cucchi2022}%
  \BibitemOpen
  \bibfield  {author} {\bibinfo {author} {\bibfnamefont {M.}~\bibnamefont
  {Cucchi}}, \bibinfo {author} {\bibfnamefont {S.}~\bibnamefont {Abreu}},
  \bibinfo {author} {\bibfnamefont {G.}~\bibnamefont {Ciccone}}, \bibinfo
  {author} {\bibfnamefont {D.}~\bibnamefont {Brunner}},\ and\ \bibinfo {author}
  {\bibfnamefont {H.}~\bibnamefont {Kleemann}},\ }\bibfield  {title} {\bibinfo
  {title} {Hands-on reservoir computing: a tutorial for practical
  implementation},\ }\href
  {https://doi.org/https://doi.org/10.3929/ethz-b-000616047} {\bibfield
  {journal} {\bibinfo  {journal} {Neuromorphic Computing and Engineering}\
  }\textbf {\bibinfo {volume} {2}},\ \bibinfo {pages} {032002} (\bibinfo {year}
  {2022})}\BibitemShut {NoStop}%
\bibitem [{\citenamefont {Kudithipudi}\ \emph {et~al.}(2016)\citenamefont
  {Kudithipudi}, \citenamefont {Saleh}, \citenamefont {Merkel}, \citenamefont
  {Thesing},\ and\ \citenamefont {Wysocki}}]{Kudithipudi2016}%
  \BibitemOpen
  \bibfield  {author} {\bibinfo {author} {\bibfnamefont {D.}~\bibnamefont
  {Kudithipudi}}, \bibinfo {author} {\bibfnamefont {Q.}~\bibnamefont {Saleh}},
  \bibinfo {author} {\bibfnamefont {C.}~\bibnamefont {Merkel}}, \bibinfo
  {author} {\bibfnamefont {J.}~\bibnamefont {Thesing}},\ and\ \bibinfo {author}
  {\bibfnamefont {B.}~\bibnamefont {Wysocki}},\ }\bibfield  {title} {\bibinfo
  {title} {Design and analysis of a neuromemristive reservoir computing
  architecture for biosignal processing},\ }\href
  {https://doi.org/https://doi.org/10.3389/fnins.2015.00502} {\bibfield
  {journal} {\bibinfo  {journal} {Frontiers in Neuroscience}\ }\textbf
  {\bibinfo {volume} {9}},\ \bibinfo {pages} {502} (\bibinfo {year}
  {2016})}\BibitemShut {NoStop}%
\bibitem [{\citenamefont {Yi}\ \emph {et~al.}(2016)\citenamefont {Yi},
  \citenamefont {Liao}, \citenamefont {Wang}, \citenamefont {Fu}, \citenamefont
  {Shen}, \citenamefont {Hou},\ and\ \citenamefont {Liu}}]{Yi2016}%
  \BibitemOpen
  \bibfield  {author} {\bibinfo {author} {\bibfnamefont {Y.}~\bibnamefont
  {Yi}}, \bibinfo {author} {\bibfnamefont {Y.}~\bibnamefont {Liao}}, \bibinfo
  {author} {\bibfnamefont {B.}~\bibnamefont {Wang}}, \bibinfo {author}
  {\bibfnamefont {X.}~\bibnamefont {Fu}}, \bibinfo {author} {\bibfnamefont
  {F.}~\bibnamefont {Shen}}, \bibinfo {author} {\bibfnamefont {H.}~\bibnamefont
  {Hou}},\ and\ \bibinfo {author} {\bibfnamefont {L.}~\bibnamefont {Liu}},\
  }\bibfield  {title} {\bibinfo {title} {{FPGA based spike-time dependent
  encoder and reservoir design in neuromorphic computing processors}},\ }\href
  {https://doi.org/10.1016/j.micpro.2016.03.009} {\bibfield  {journal}
  {\bibinfo  {journal} {Microprocessors and Microsystems}\ }\textbf {\bibinfo
  {volume} {46}},\ \bibinfo {pages} {175} (\bibinfo {year} {2016})}\BibitemShut
  {NoStop}%
\bibitem [{\citenamefont {Tanaka}\ \emph {et~al.}(2019)\citenamefont {Tanaka},
  \citenamefont {Yamane}, \citenamefont {H\'{e}roux}, \citenamefont {Nakane},
  \citenamefont {Kanazawa}, \citenamefont {Takeda}, \citenamefont {Numata},
  \citenamefont {Nakano},\ and\ \citenamefont {Hirose}}]{Tanaka2019}%
  \BibitemOpen
  \bibfield  {author} {\bibinfo {author} {\bibfnamefont {G.}~\bibnamefont
  {Tanaka}}, \bibinfo {author} {\bibfnamefont {T.}~\bibnamefont {Yamane}},
  \bibinfo {author} {\bibfnamefont {J.~B.}\ \bibnamefont {H\'{e}roux}},
  \bibinfo {author} {\bibfnamefont {R.}~\bibnamefont {Nakane}}, \bibinfo
  {author} {\bibfnamefont {N.}~\bibnamefont {Kanazawa}}, \bibinfo {author}
  {\bibfnamefont {S.}~\bibnamefont {Takeda}}, \bibinfo {author} {\bibfnamefont
  {H.}~\bibnamefont {Numata}}, \bibinfo {author} {\bibfnamefont
  {D.}~\bibnamefont {Nakano}},\ and\ \bibinfo {author} {\bibfnamefont
  {A.}~\bibnamefont {Hirose}},\ }\bibfield  {title} {\bibinfo {title} {{Recent
  advances in physical reservoir computing: A review}},\ }\href
  {https://doi.org/10.1016/j.neunet.2019.03.005} {\bibfield  {journal}
  {\bibinfo  {journal} {Neural Networks}\ }\textbf {\bibinfo {volume} {115}},\
  \bibinfo {pages} {100} (\bibinfo {year} {2019})}\BibitemShut {NoStop}%
\bibitem [{\citenamefont {Nakajima}(2020)}]{Nakajima2020}%
  \BibitemOpen
  \bibfield  {author} {\bibinfo {author} {\bibfnamefont {K.}~\bibnamefont
  {Nakajima}},\ }\bibfield  {title} {\bibinfo {title} {Physical reservoir
  computing—an introductory perspective},\ }\href
  {https://doi.org/10.35848/1347-4065/ab8d4f} {\bibfield  {journal} {\bibinfo
  {journal} {Japanese Journal of Applied Physics}\ }\textbf {\bibinfo {volume}
  {59}},\ \bibinfo {pages} {060501} (\bibinfo {year} {2020})}\BibitemShut
  {NoStop}%
\bibitem [{\citenamefont {Christensen}\ \emph {et~al.}(2022)\citenamefont
  {Christensen}, \citenamefont {Dittmann}, \citenamefont {Linares-Barranco},
  \citenamefont {Sebastian}, \citenamefont {Le~Gallo}, \citenamefont
  {Redaelli}, \citenamefont {Slesazeck}, \citenamefont {Mikolajick},
  \citenamefont {Spiga}, \citenamefont {Menzel}, \citenamefont {Valov},
  \citenamefont {Milano}, \citenamefont {Ricciardi}, \citenamefont {Liang},
  \citenamefont {Miao}, \citenamefont {Lanza}, \citenamefont {Quill},
  \citenamefont {Keene}, \citenamefont {Salleo}, \citenamefont {Grollier},
  \citenamefont {Marković}, \citenamefont {Mizrahi}, \citenamefont {Yao},
  \citenamefont {Yang}, \citenamefont {Indiveri}, \citenamefont {Strachan},
  \citenamefont {Datta}, \citenamefont {Vianello}, \citenamefont {Valentian},
  \citenamefont {Feldmann}, \citenamefont {Li}, \citenamefont {Pernice},
  \citenamefont {Bhaskaran}, \citenamefont {Furber}, \citenamefont {Neftci},
  \citenamefont {Scherr}, \citenamefont {Maass}, \citenamefont {Ramaswamy},
  \citenamefont {Tapson}, \citenamefont {Panda}, \citenamefont {Kim},
  \citenamefont {Tanaka}, \citenamefont {Thorpe}, \citenamefont {Bartolozzi},
  \citenamefont {Cleland}, \citenamefont {Posch}, \citenamefont {Liu},
  \citenamefont {Panuccio}, \citenamefont {Mahmud}, \citenamefont {Mazumder},
  \citenamefont {Hosseini}, \citenamefont {Mohsenin}, \citenamefont {Donati},
  \citenamefont {Tolu}, \citenamefont {Galeazzi}, \citenamefont {Christensen},
  \citenamefont {Holm}, \citenamefont {Ielmini},\ and\ \citenamefont
  {Pryds}}]{Christensen2022}%
  \BibitemOpen
  \bibfield  {author} {\bibinfo {author} {\bibfnamefont {D.~V.}\ \bibnamefont
  {Christensen}}, \bibinfo {author} {\bibfnamefont {R.}~\bibnamefont
  {Dittmann}}, \bibinfo {author} {\bibfnamefont {B.}~\bibnamefont
  {Linares-Barranco}}, \bibinfo {author} {\bibfnamefont {A.}~\bibnamefont
  {Sebastian}}, \bibinfo {author} {\bibfnamefont {M.}~\bibnamefont {Le~Gallo}},
  \bibinfo {author} {\bibfnamefont {A.}~\bibnamefont {Redaelli}}, \bibinfo
  {author} {\bibfnamefont {S.}~\bibnamefont {Slesazeck}}, \bibinfo {author}
  {\bibfnamefont {T.}~\bibnamefont {Mikolajick}}, \bibinfo {author}
  {\bibfnamefont {S.}~\bibnamefont {Spiga}}, \bibinfo {author} {\bibfnamefont
  {S.}~\bibnamefont {Menzel}}, \bibinfo {author} {\bibfnamefont
  {I.}~\bibnamefont {Valov}}, \bibinfo {author} {\bibfnamefont
  {G.}~\bibnamefont {Milano}}, \bibinfo {author} {\bibfnamefont
  {C.}~\bibnamefont {Ricciardi}}, \bibinfo {author} {\bibfnamefont {S.-J.}\
  \bibnamefont {Liang}}, \bibinfo {author} {\bibfnamefont {F.}~\bibnamefont
  {Miao}}, \bibinfo {author} {\bibfnamefont {M.}~\bibnamefont {Lanza}},
  \bibinfo {author} {\bibfnamefont {T.~J.}\ \bibnamefont {Quill}}, \bibinfo
  {author} {\bibfnamefont {S.~T.}\ \bibnamefont {Keene}}, \bibinfo {author}
  {\bibfnamefont {A.}~\bibnamefont {Salleo}}, \bibinfo {author} {\bibfnamefont
  {J.}~\bibnamefont {Grollier}}, \bibinfo {author} {\bibfnamefont
  {D.}~\bibnamefont {Marković}}, \bibinfo {author} {\bibfnamefont
  {A.}~\bibnamefont {Mizrahi}}, \bibinfo {author} {\bibfnamefont
  {P.}~\bibnamefont {Yao}}, \bibinfo {author} {\bibfnamefont {J.~J.}\
  \bibnamefont {Yang}}, \bibinfo {author} {\bibfnamefont {G.}~\bibnamefont
  {Indiveri}}, \bibinfo {author} {\bibfnamefont {J.~P.}\ \bibnamefont
  {Strachan}}, \bibinfo {author} {\bibfnamefont {S.}~\bibnamefont {Datta}},
  \bibinfo {author} {\bibfnamefont {E.}~\bibnamefont {Vianello}}, \bibinfo
  {author} {\bibfnamefont {A.}~\bibnamefont {Valentian}}, \bibinfo {author}
  {\bibfnamefont {J.}~\bibnamefont {Feldmann}}, \bibinfo {author}
  {\bibfnamefont {X.}~\bibnamefont {Li}}, \bibinfo {author} {\bibfnamefont
  {W.~H.~P.}\ \bibnamefont {Pernice}}, \bibinfo {author} {\bibfnamefont
  {H.}~\bibnamefont {Bhaskaran}}, \bibinfo {author} {\bibfnamefont
  {S.}~\bibnamefont {Furber}}, \bibinfo {author} {\bibfnamefont
  {E.}~\bibnamefont {Neftci}}, \bibinfo {author} {\bibfnamefont
  {F.}~\bibnamefont {Scherr}}, \bibinfo {author} {\bibfnamefont
  {W.}~\bibnamefont {Maass}}, \bibinfo {author} {\bibfnamefont
  {S.}~\bibnamefont {Ramaswamy}}, \bibinfo {author} {\bibfnamefont
  {J.}~\bibnamefont {Tapson}}, \bibinfo {author} {\bibfnamefont
  {P.}~\bibnamefont {Panda}}, \bibinfo {author} {\bibfnamefont
  {Y.}~\bibnamefont {Kim}}, \bibinfo {author} {\bibfnamefont {G.}~\bibnamefont
  {Tanaka}}, \bibinfo {author} {\bibfnamefont {S.}~\bibnamefont {Thorpe}},
  \bibinfo {author} {\bibfnamefont {C.}~\bibnamefont {Bartolozzi}}, \bibinfo
  {author} {\bibfnamefont {T.~A.}\ \bibnamefont {Cleland}}, \bibinfo {author}
  {\bibfnamefont {C.}~\bibnamefont {Posch}}, \bibinfo {author} {\bibfnamefont
  {S.}~\bibnamefont {Liu}}, \bibinfo {author} {\bibfnamefont {G.}~\bibnamefont
  {Panuccio}}, \bibinfo {author} {\bibfnamefont {M.}~\bibnamefont {Mahmud}},
  \bibinfo {author} {\bibfnamefont {A.~N.}\ \bibnamefont {Mazumder}}, \bibinfo
  {author} {\bibfnamefont {M.}~\bibnamefont {Hosseini}}, \bibinfo {author}
  {\bibfnamefont {T.}~\bibnamefont {Mohsenin}}, \bibinfo {author}
  {\bibfnamefont {E.}~\bibnamefont {Donati}}, \bibinfo {author} {\bibfnamefont
  {S.}~\bibnamefont {Tolu}}, \bibinfo {author} {\bibfnamefont {R.}~\bibnamefont
  {Galeazzi}}, \bibinfo {author} {\bibfnamefont {M.~E.}\ \bibnamefont
  {Christensen}}, \bibinfo {author} {\bibfnamefont {S.}~\bibnamefont {Holm}},
  \bibinfo {author} {\bibfnamefont {D.}~\bibnamefont {Ielmini}},\ and\ \bibinfo
  {author} {\bibfnamefont {N.}~\bibnamefont {Pryds}},\ }\bibfield  {title}
  {\bibinfo {title} {2022 roadmap on neuromorphic computing and engineering},\
  }\href {https://doi.org/10.1088/2634-4386/ac4a83} {\bibfield  {journal}
  {\bibinfo  {journal} {Neuromorphic Computing and Engineering}\ }\textbf
  {\bibinfo {volume} {2}},\ \bibinfo {pages} {022501} (\bibinfo {year}
  {2022})}\BibitemShut {NoStop}%
\bibitem [{\citenamefont {Grollier}\ \emph {et~al.}(2020)\citenamefont
  {Grollier}, \citenamefont {Querlioz}, \citenamefont {Camsari}, \citenamefont
  {Everschor-Sitte}, \citenamefont {Fukami},\ and\ \citenamefont
  {Stiles}}]{Grollier2020}%
  \BibitemOpen
  \bibfield  {author} {\bibinfo {author} {\bibfnamefont {J.}~\bibnamefont
  {Grollier}}, \bibinfo {author} {\bibfnamefont {D.}~\bibnamefont {Querlioz}},
  \bibinfo {author} {\bibfnamefont {K.}~\bibnamefont {Camsari}}, \bibinfo
  {author} {\bibfnamefont {K.}~\bibnamefont {Everschor-Sitte}}, \bibinfo
  {author} {\bibfnamefont {S.}~\bibnamefont {Fukami}},\ and\ \bibinfo {author}
  {\bibfnamefont {M.~D.}\ \bibnamefont {Stiles}},\ }\bibfield  {title}
  {\bibinfo {title} {Neuromorphic spintronics},\ }\href
  {https://doi.org/10.1038/s41928-019-0360-9} {\bibfield  {journal} {\bibinfo
  {journal} {Nature electronics}\ }\textbf {\bibinfo {volume} {3}},\ \bibinfo
  {pages} {360} (\bibinfo {year} {2020})}\BibitemShut {NoStop}%
\bibitem [{\citenamefont {Roede}\ \emph {et~al.}(2022)\citenamefont {Roede},
  \citenamefont {Shapovalov}, \citenamefont {Moran}, \citenamefont {Mosberg},
  \citenamefont {Yan}, \citenamefont {Bourret}, \citenamefont {Cano},
  \citenamefont {Huey}, \citenamefont {van Helvoort},\ and\ \citenamefont
  {Meier}}]{Roede2022}%
  \BibitemOpen
  \bibfield  {author} {\bibinfo {author} {\bibfnamefont {E.~D.}\ \bibnamefont
  {Roede}}, \bibinfo {author} {\bibfnamefont {K.}~\bibnamefont {Shapovalov}},
  \bibinfo {author} {\bibfnamefont {T.~J.}\ \bibnamefont {Moran}}, \bibinfo
  {author} {\bibfnamefont {A.~B.}\ \bibnamefont {Mosberg}}, \bibinfo {author}
  {\bibfnamefont {Z.}~\bibnamefont {Yan}}, \bibinfo {author} {\bibfnamefont
  {E.}~\bibnamefont {Bourret}}, \bibinfo {author} {\bibfnamefont
  {A.}~\bibnamefont {Cano}}, \bibinfo {author} {\bibfnamefont {B.~D.}\
  \bibnamefont {Huey}}, \bibinfo {author} {\bibfnamefont {A.~T.~J.}\
  \bibnamefont {van Helvoort}},\ and\ \bibinfo {author} {\bibfnamefont
  {D.}~\bibnamefont {Meier}},\ }\bibfield  {title} {\bibinfo {title} {The third
  dimension of ferroelectric domain walls},\ }\href
  {https://doi.org/10.1002/adma.202202614} {\bibfield  {journal} {\bibinfo
  {journal} {Advanced Materials}\ }\textbf {\bibinfo {volume} {34}},\ \bibinfo
  {pages} {2202614} (\bibinfo {year} {2022})}\BibitemShut {NoStop}%
\bibitem [{\citenamefont {Furuta}\ \emph {et~al.}(2018)\citenamefont {Furuta},
  \citenamefont {Fujii}, \citenamefont {Nakajima}, \citenamefont {Tsunegi},
  \citenamefont {Kubota}, \citenamefont {Suzuki},\ and\ \citenamefont
  {Miwa}}]{Furuta2018}%
  \BibitemOpen
  \bibfield  {author} {\bibinfo {author} {\bibfnamefont {T.}~\bibnamefont
  {Furuta}}, \bibinfo {author} {\bibfnamefont {K.}~\bibnamefont {Fujii}},
  \bibinfo {author} {\bibfnamefont {K.}~\bibnamefont {Nakajima}}, \bibinfo
  {author} {\bibfnamefont {S.}~\bibnamefont {Tsunegi}}, \bibinfo {author}
  {\bibfnamefont {H.}~\bibnamefont {Kubota}}, \bibinfo {author} {\bibfnamefont
  {Y.}~\bibnamefont {Suzuki}},\ and\ \bibinfo {author} {\bibfnamefont
  {S.}~\bibnamefont {Miwa}},\ }\bibfield  {title} {\bibinfo {title}
  {{Macromagnetic Simulation for Reservoir Computing Utilizing Spin Dynamics in
  Magnetic Tunnel Junctions}},\ }\href
  {https://doi.org/10.1103/PhysRevApplied.10.034063} {\bibfield  {journal}
  {\bibinfo  {journal} {Physical Review Applied}\ }\textbf {\bibinfo {volume}
  {10}},\ \bibinfo {pages} {034063} (\bibinfo {year} {2018})}\BibitemShut
  {NoStop}%
\bibitem [{\citenamefont {Torrejon}\ \emph {et~al.}(2017)\citenamefont
  {Torrejon}, \citenamefont {Riou}, \citenamefont {Araujo}, \citenamefont
  {Tsunegi}, \citenamefont {Khalsa}, \citenamefont {Querlioz}, \citenamefont
  {Bortolotti}, \citenamefont {Cros}, \citenamefont {Yakushiji}, \citenamefont
  {Fukushima}, \citenamefont {Kubota}, \citenamefont {Yuasa}, \citenamefont
  {Stiles},\ and\ \citenamefont {Grollier}}]{Torrejon2017}%
  \BibitemOpen
  \bibfield  {author} {\bibinfo {author} {\bibfnamefont {J.}~\bibnamefont
  {Torrejon}}, \bibinfo {author} {\bibfnamefont {M.}~\bibnamefont {Riou}},
  \bibinfo {author} {\bibfnamefont {F.~A.}\ \bibnamefont {Araujo}}, \bibinfo
  {author} {\bibfnamefont {S.}~\bibnamefont {Tsunegi}}, \bibinfo {author}
  {\bibfnamefont {G.}~\bibnamefont {Khalsa}}, \bibinfo {author} {\bibfnamefont
  {D.}~\bibnamefont {Querlioz}}, \bibinfo {author} {\bibfnamefont
  {P.}~\bibnamefont {Bortolotti}}, \bibinfo {author} {\bibfnamefont
  {V.}~\bibnamefont {Cros}}, \bibinfo {author} {\bibfnamefont {K.}~\bibnamefont
  {Yakushiji}}, \bibinfo {author} {\bibfnamefont {A.}~\bibnamefont
  {Fukushima}}, \bibinfo {author} {\bibfnamefont {H.}~\bibnamefont {Kubota}},
  \bibinfo {author} {\bibfnamefont {S.}~\bibnamefont {Yuasa}}, \bibinfo
  {author} {\bibfnamefont {M.~D.}\ \bibnamefont {Stiles}},\ and\ \bibinfo
  {author} {\bibfnamefont {J.}~\bibnamefont {Grollier}},\ }\bibfield  {title}
  {\bibinfo {title} {{Neuromorphic computing with nanoscale spintronic
  oscillators}},\ }\href {https://doi.org/10.1038/nature23011} {\bibfield
  {journal} {\bibinfo  {journal} {Nature}\ }\textbf {\bibinfo {volume} {547}},\
  \bibinfo {pages} {428} (\bibinfo {year} {2017})}\BibitemShut {NoStop}%
\bibitem [{\citenamefont {Marković}\ \emph {et~al.}(2019)\citenamefont
  {Marković}, \citenamefont {Leroux}, \citenamefont {Riou}, \citenamefont
  {Abreu~Araujo}, \citenamefont {Torrejon}, \citenamefont {Querlioz},
  \citenamefont {Fukushima}, \citenamefont {Yuasa}, \citenamefont {Trastoy},
  \citenamefont {Bortolotti},\ and\ \citenamefont {Grollier}}]{Markovic2019}%
  \BibitemOpen
  \bibfield  {author} {\bibinfo {author} {\bibfnamefont {D.}~\bibnamefont
  {Marković}}, \bibinfo {author} {\bibfnamefont {N.}~\bibnamefont {Leroux}},
  \bibinfo {author} {\bibfnamefont {M.}~\bibnamefont {Riou}}, \bibinfo {author}
  {\bibfnamefont {F.}~\bibnamefont {Abreu~Araujo}}, \bibinfo {author}
  {\bibfnamefont {J.}~\bibnamefont {Torrejon}}, \bibinfo {author}
  {\bibfnamefont {D.}~\bibnamefont {Querlioz}}, \bibinfo {author}
  {\bibfnamefont {A.}~\bibnamefont {Fukushima}}, \bibinfo {author}
  {\bibfnamefont {S.}~\bibnamefont {Yuasa}}, \bibinfo {author} {\bibfnamefont
  {J.}~\bibnamefont {Trastoy}}, \bibinfo {author} {\bibfnamefont
  {P.}~\bibnamefont {Bortolotti}},\ and\ \bibinfo {author} {\bibfnamefont
  {J.}~\bibnamefont {Grollier}},\ }\bibfield  {title} {\bibinfo {title}
  {{Reservoir computing with the frequency, phase, and amplitude of spin-torque
  nano-oscillators}},\ }\href {https://doi.org/10.1063/1.5079305} {\bibfield
  {journal} {\bibinfo  {journal} {Applied Physics Letters}\ }\textbf {\bibinfo
  {volume} {114}},\ \bibinfo {pages} {012409} (\bibinfo {year}
  {2019})}\BibitemShut {NoStop}%
\bibitem [{\citenamefont {Van~der Sande}\ \emph {et~al.}(2017)\citenamefont
  {Van~der Sande}, \citenamefont {Brunner},\ and\ \citenamefont
  {Soriano}}]{Van2017}%
  \BibitemOpen
  \bibfield  {author} {\bibinfo {author} {\bibfnamefont {G.}~\bibnamefont
  {Van~der Sande}}, \bibinfo {author} {\bibfnamefont {D.}~\bibnamefont
  {Brunner}},\ and\ \bibinfo {author} {\bibfnamefont {M.~C.}\ \bibnamefont
  {Soriano}},\ }\bibfield  {title} {\bibinfo {title} {Advances in photonic
  reservoir computing},\ }\href {https://doi.org/10.1515/nanoph-2016-0132}
  {\bibfield  {journal} {\bibinfo  {journal} {Nanophotonics}\ }\textbf
  {\bibinfo {volume} {6}},\ \bibinfo {pages} {561} (\bibinfo {year}
  {2017})}\BibitemShut {NoStop}%
\bibitem [{\citenamefont {R{\"o}hm}\ and\ \citenamefont
  {L{\"u}dge}(2018)}]{Rohm2018}%
  \BibitemOpen
  \bibfield  {author} {\bibinfo {author} {\bibfnamefont {A.}~\bibnamefont
  {R{\"o}hm}}\ and\ \bibinfo {author} {\bibfnamefont {K.}~\bibnamefont
  {L{\"u}dge}},\ }\bibfield  {title} {\bibinfo {title} {Multiplexed networks:
  reservoir computing with virtual and real nodes},\ }\href
  {https://doi.org/10.1088/2399-6528/aad56d} {\bibfield  {journal} {\bibinfo
  {journal} {Journal of Physics Communications}\ }\textbf {\bibinfo {volume}
  {2}},\ \bibinfo {pages} {085007} (\bibinfo {year} {2018})}\BibitemShut
  {NoStop}%
\bibitem [{\citenamefont {Nomura}\ \emph {et~al.}(2019)\citenamefont {Nomura},
  \citenamefont {Furuta}, \citenamefont {Tsujimoto}, \citenamefont
  {Kuwabiraki}, \citenamefont {Peper}, \citenamefont {Tamura}, \citenamefont
  {Miwa}, \citenamefont {Goto}, \citenamefont {Nakatani},\ and\ \citenamefont
  {Suzuki}}]{Nomura2019}%
  \BibitemOpen
  \bibfield  {author} {\bibinfo {author} {\bibfnamefont {H.}~\bibnamefont
  {Nomura}}, \bibinfo {author} {\bibfnamefont {T.}~\bibnamefont {Furuta}},
  \bibinfo {author} {\bibfnamefont {K.}~\bibnamefont {Tsujimoto}}, \bibinfo
  {author} {\bibfnamefont {Y.}~\bibnamefont {Kuwabiraki}}, \bibinfo {author}
  {\bibfnamefont {F.}~\bibnamefont {Peper}}, \bibinfo {author} {\bibfnamefont
  {E.}~\bibnamefont {Tamura}}, \bibinfo {author} {\bibfnamefont
  {S.}~\bibnamefont {Miwa}}, \bibinfo {author} {\bibfnamefont {M.}~\bibnamefont
  {Goto}}, \bibinfo {author} {\bibfnamefont {R.}~\bibnamefont {Nakatani}},\
  and\ \bibinfo {author} {\bibfnamefont {Y.}~\bibnamefont {Suzuki}},\
  }\bibfield  {title} {\bibinfo {title} {{Reservoir computing with
  dipole-coupled nanomagnets}},\ }\href
  {https://doi.org/https://iopscience.iop.org/article/10.7567/1347-4065/ab2406}
  {\bibfield  {journal} {\bibinfo  {journal} {Japanese Journal of Applied
  Physics}\ }\textbf {\bibinfo {volume} {58}},\ \bibinfo {pages} {070901}
  (\bibinfo {year} {2019})}\BibitemShut {NoStop}%
\bibitem [{\citenamefont {Gartside}\ \emph {et~al.}(2022)\citenamefont
  {Gartside}, \citenamefont {Stenning}, \citenamefont {Vanstone}, \citenamefont
  {Holder}, \citenamefont {Arroo}, \citenamefont {Dion}, \citenamefont
  {Caravelli}, \citenamefont {Kurebayashi},\ and\ \citenamefont
  {Branford}}]{Gartside2022}%
  \BibitemOpen
  \bibfield  {author} {\bibinfo {author} {\bibfnamefont {J.~C.}\ \bibnamefont
  {Gartside}}, \bibinfo {author} {\bibfnamefont {K.~D.}\ \bibnamefont
  {Stenning}}, \bibinfo {author} {\bibfnamefont {A.}~\bibnamefont {Vanstone}},
  \bibinfo {author} {\bibfnamefont {H.~H.}\ \bibnamefont {Holder}}, \bibinfo
  {author} {\bibfnamefont {D.~M.}\ \bibnamefont {Arroo}}, \bibinfo {author}
  {\bibfnamefont {T.}~\bibnamefont {Dion}}, \bibinfo {author} {\bibfnamefont
  {F.}~\bibnamefont {Caravelli}}, \bibinfo {author} {\bibfnamefont
  {H.}~\bibnamefont {Kurebayashi}},\ and\ \bibinfo {author} {\bibfnamefont
  {W.~R.}\ \bibnamefont {Branford}},\ }\bibfield  {title} {\bibinfo {title}
  {Reconfigurable training and reservoir computing in an artificial spin-vortex
  ice via spin-wave fingerprinting},\ }\href
  {https://doi.org/10.1038/s41565-022-01091-7} {\bibfield  {journal} {\bibinfo
  {journal} {Nature Nanotechnology}\ }\textbf {\bibinfo {volume} {17}},\
  \bibinfo {pages} {460} (\bibinfo {year} {2022})}\BibitemShut {NoStop}%
\bibitem [{\citenamefont {Vidamour}\ \emph {et~al.}(2023)\citenamefont
  {Vidamour}, \citenamefont {Swindells}, \citenamefont {Venkat}, \citenamefont
  {Manneschi}, \citenamefont {Fry}, \citenamefont {Welbourne}, \citenamefont
  {Rowan-Robinson}, \citenamefont {Backes}, \citenamefont {Maccherozzi},
  \citenamefont {Dhesi}, \citenamefont {Vasilaki}, \citenamefont {Allwood},\
  and\ \citenamefont {Hayward}}]{Vidamour2023}%
  \BibitemOpen
  \bibfield  {author} {\bibinfo {author} {\bibfnamefont {I.~T.}\ \bibnamefont
  {Vidamour}}, \bibinfo {author} {\bibfnamefont {C.}~\bibnamefont {Swindells}},
  \bibinfo {author} {\bibfnamefont {G.}~\bibnamefont {Venkat}}, \bibinfo
  {author} {\bibfnamefont {L.}~\bibnamefont {Manneschi}}, \bibinfo {author}
  {\bibfnamefont {P.~W.}\ \bibnamefont {Fry}}, \bibinfo {author} {\bibfnamefont
  {A.}~\bibnamefont {Welbourne}}, \bibinfo {author} {\bibfnamefont {R.~M.}\
  \bibnamefont {Rowan-Robinson}}, \bibinfo {author} {\bibfnamefont
  {D.}~\bibnamefont {Backes}}, \bibinfo {author} {\bibfnamefont
  {F.}~\bibnamefont {Maccherozzi}}, \bibinfo {author} {\bibfnamefont {S.~S.}\
  \bibnamefont {Dhesi}}, \bibinfo {author} {\bibfnamefont {E.}~\bibnamefont
  {Vasilaki}}, \bibinfo {author} {\bibfnamefont {D.~A.}\ \bibnamefont
  {Allwood}},\ and\ \bibinfo {author} {\bibfnamefont {T.~J.}\ \bibnamefont
  {Hayward}},\ }\bibfield  {title} {\bibinfo {title} {Reconfigurable reservoir
  computing in a magnetic metamaterial},\ }\href
  {https://doi.org/10.1038/s42005-023-01352-4} {\bibfield  {journal} {\bibinfo
  {journal} {Communications Physics}\ }\textbf {\bibinfo {volume} {6}}
  (\bibinfo {year} {2023})}\BibitemShut {NoStop}%
\bibitem [{\citenamefont {Raab}\ \emph {et~al.}(2022)\citenamefont {Raab},
  \citenamefont {Brems}, \citenamefont {Beneke}, \citenamefont {Dohi},
  \citenamefont {Roth{\"o}rl}, \citenamefont {Kammerbauer}, \citenamefont
  {Mentink},\ and\ \citenamefont {Kl{\"a}ui}}]{Raab2022}%
  \BibitemOpen
  \bibfield  {author} {\bibinfo {author} {\bibfnamefont {K.}~\bibnamefont
  {Raab}}, \bibinfo {author} {\bibfnamefont {M.~A.}\ \bibnamefont {Brems}},
  \bibinfo {author} {\bibfnamefont {G.}~\bibnamefont {Beneke}}, \bibinfo
  {author} {\bibfnamefont {T.}~\bibnamefont {Dohi}}, \bibinfo {author}
  {\bibfnamefont {J.}~\bibnamefont {Roth{\"o}rl}}, \bibinfo {author}
  {\bibfnamefont {F.}~\bibnamefont {Kammerbauer}}, \bibinfo {author}
  {\bibfnamefont {J.~H.}\ \bibnamefont {Mentink}},\ and\ \bibinfo {author}
  {\bibfnamefont {M.}~\bibnamefont {Kl{\"a}ui}},\ }\bibfield  {title} {\bibinfo
  {title} {Brownian reservoir computing realized using geometrically confined
  skyrmion dynamics},\ }\href {https://doi.org/10.1038/s41467-022-34309-2}
  {\bibfield  {journal} {\bibinfo  {journal} {Nature Communications}\ }\textbf
  {\bibinfo {volume} {13}},\ \bibinfo {pages} {6982} (\bibinfo {year}
  {2022})}\BibitemShut {NoStop}%
\bibitem [{\citenamefont {Prychynenko}\ \emph {et~al.}(2018)\citenamefont
  {Prychynenko}, \citenamefont {Sitte}, \citenamefont {Litzius}, \citenamefont
  {Kr\"uger}, \citenamefont {Bourianoff}, \citenamefont {Kl\"aui},
  \citenamefont {Sinova},\ and\ \citenamefont
  {Everschor-Sitte}}]{Prychynenko2018}%
  \BibitemOpen
  \bibfield  {author} {\bibinfo {author} {\bibfnamefont {D.}~\bibnamefont
  {Prychynenko}}, \bibinfo {author} {\bibfnamefont {M.}~\bibnamefont {Sitte}},
  \bibinfo {author} {\bibfnamefont {K.}~\bibnamefont {Litzius}}, \bibinfo
  {author} {\bibfnamefont {B.}~\bibnamefont {Kr\"uger}}, \bibinfo {author}
  {\bibfnamefont {G.}~\bibnamefont {Bourianoff}}, \bibinfo {author}
  {\bibfnamefont {M.}~\bibnamefont {Kl\"aui}}, \bibinfo {author} {\bibfnamefont
  {J.}~\bibnamefont {Sinova}},\ and\ \bibinfo {author} {\bibfnamefont
  {K.}~\bibnamefont {Everschor-Sitte}},\ }\bibfield  {title} {\bibinfo {title}
  {{Magnetic Skyrmion as a Nonlinear Resistive Element: A Potential Building
  Block for Reservoir Computing}},\ }\href
  {https://doi.org/10.1103/PhysRevApplied.9.014034} {\bibfield  {journal}
  {\bibinfo  {journal} {Phys. Rev. Applied}\ }\textbf {\bibinfo {volume} {9}},\
  \bibinfo {pages} {014034} (\bibinfo {year} {2018})}\BibitemShut {NoStop}%
\bibitem [{\citenamefont {Bourianoff}\ \emph {et~al.}(2018)\citenamefont
  {Bourianoff}, \citenamefont {Pinna}, \citenamefont {Sitte},\ and\
  \citenamefont {Everschor-Sitte}}]{Bourianoff2018}%
  \BibitemOpen
  \bibfield  {author} {\bibinfo {author} {\bibfnamefont {G.}~\bibnamefont
  {Bourianoff}}, \bibinfo {author} {\bibfnamefont {D.}~\bibnamefont {Pinna}},
  \bibinfo {author} {\bibfnamefont {M.}~\bibnamefont {Sitte}},\ and\ \bibinfo
  {author} {\bibfnamefont {K.}~\bibnamefont {Everschor-Sitte}},\ }\bibfield
  {title} {\bibinfo {title} {{Potential implementation of reservoir computing
  models based on magnetic skyrmions}},\ }\href
  {https://doi.org/10.1063/1.5006918} {\bibfield  {journal} {\bibinfo
  {journal} {AIP Advances}\ }\textbf {\bibinfo {volume} {8}},\ \bibinfo {pages}
  {055602} (\bibinfo {year} {2018})}\BibitemShut {NoStop}%
\bibitem [{\citenamefont {Pinna}\ \emph {et~al.}(2020)\citenamefont {Pinna},
  \citenamefont {Bourianoff},\ and\ \citenamefont
  {Everschor-Sitte}}]{Pinna2018}%
  \BibitemOpen
  \bibfield  {author} {\bibinfo {author} {\bibfnamefont {D.}~\bibnamefont
  {Pinna}}, \bibinfo {author} {\bibfnamefont {G.}~\bibnamefont {Bourianoff}},\
  and\ \bibinfo {author} {\bibfnamefont {K.}~\bibnamefont {Everschor-Sitte}},\
  }\bibfield  {title} {\bibinfo {title} {{Reservoir Computing with Random
  Skyrmion Textures}},\ }\href
  {https://doi.org/10.1103/physrevapplied.14.054020} {\bibfield  {journal}
  {\bibinfo  {journal} {Physical Review Applied}\ }\textbf {\bibinfo {volume}
  {14}},\ \bibinfo {pages} {054020} (\bibinfo {year} {2020})}\BibitemShut
  {NoStop}%
\bibitem [{\citenamefont {Sun}\ \emph {et~al.}(2023)\citenamefont {Sun},
  \citenamefont {Lin}, \citenamefont {Lei}, \citenamefont {Chen}, \citenamefont
  {Kang}, \citenamefont {Zhao}, \citenamefont {Wei}, \citenamefont {Chen},
  \citenamefont {Pang}, \citenamefont {Hu}, \citenamefont {Yang}, \citenamefont
  {Dong}, \citenamefont {Zhao}, \citenamefont {Liu}, \citenamefont {Yuan},
  \citenamefont {Ullrich}, \citenamefont {Back}, \citenamefont {Zhang},
  \citenamefont {Pan}, \citenamefont {Zhao}, \citenamefont {Feng},
  \citenamefont {Fert},\ and\ \citenamefont {Zhao}}]{Sun2023}%
  \BibitemOpen
  \bibfield  {author} {\bibinfo {author} {\bibfnamefont {Y.}~\bibnamefont
  {Sun}}, \bibinfo {author} {\bibfnamefont {T.}~\bibnamefont {Lin}}, \bibinfo
  {author} {\bibfnamefont {N.}~\bibnamefont {Lei}}, \bibinfo {author}
  {\bibfnamefont {X.}~\bibnamefont {Chen}}, \bibinfo {author} {\bibfnamefont
  {W.}~\bibnamefont {Kang}}, \bibinfo {author} {\bibfnamefont {Z.}~\bibnamefont
  {Zhao}}, \bibinfo {author} {\bibfnamefont {D.}~\bibnamefont {Wei}}, \bibinfo
  {author} {\bibfnamefont {C.}~\bibnamefont {Chen}}, \bibinfo {author}
  {\bibfnamefont {S.}~\bibnamefont {Pang}}, \bibinfo {author} {\bibfnamefont
  {L.}~\bibnamefont {Hu}}, \bibinfo {author} {\bibfnamefont {L.}~\bibnamefont
  {Yang}}, \bibinfo {author} {\bibfnamefont {E.}~\bibnamefont {Dong}}, \bibinfo
  {author} {\bibfnamefont {L.}~\bibnamefont {Zhao}}, \bibinfo {author}
  {\bibfnamefont {L.}~\bibnamefont {Liu}}, \bibinfo {author} {\bibfnamefont
  {Z.}~\bibnamefont {Yuan}}, \bibinfo {author} {\bibfnamefont {A.}~\bibnamefont
  {Ullrich}}, \bibinfo {author} {\bibfnamefont {C.~H.}\ \bibnamefont {Back}},
  \bibinfo {author} {\bibfnamefont {J.}~\bibnamefont {Zhang}}, \bibinfo
  {author} {\bibfnamefont {D.}~\bibnamefont {Pan}}, \bibinfo {author}
  {\bibfnamefont {J.}~\bibnamefont {Zhao}}, \bibinfo {author} {\bibfnamefont
  {M.}~\bibnamefont {Feng}}, \bibinfo {author} {\bibfnamefont {A.}~\bibnamefont
  {Fert}},\ and\ \bibinfo {author} {\bibfnamefont {W.}~\bibnamefont {Zhao}},\
  }\bibfield  {title} {\bibinfo {title} {Experimental demonstration of a
  skyrmion-enhanced strain-mediated physical reservoir computing system},\
  }\href {https://doi.org/https://www.nature.com/articles/s41467-023-39207-9}
  {\bibfield  {journal} {\bibinfo  {journal} {Nature Communications}\ }\textbf
  {\bibinfo {volume} {14}},\ \bibinfo {pages} {3434} (\bibinfo {year}
  {2023})}\BibitemShut {NoStop}%
\bibitem [{\citenamefont {Lee}\ and\ \citenamefont
  {Mochizuki}(2022)}]{LeeMK2022}%
  \BibitemOpen
  \bibfield  {author} {\bibinfo {author} {\bibfnamefont {M.-K.}\ \bibnamefont
  {Lee}}\ and\ \bibinfo {author} {\bibfnamefont {M.}~\bibnamefont
  {Mochizuki}},\ }\bibfield  {title} {\bibinfo {title} {Reservoir computing
  with spin waves in a skyrmion crystal},\ }\href
  {https://doi.org/10.1103/PhysRevApplied.18.014074} {\bibfield  {journal}
  {\bibinfo  {journal} {Phys. Rev. Appl.}\ }\textbf {\bibinfo {volume} {18}},\
  \bibinfo {pages} {014074} (\bibinfo {year} {2022})}\BibitemShut {NoStop}%
\bibitem [{\citenamefont {Lee}\ and\ \citenamefont
  {Mochizuki}(2023)}]{LeeMK2023}%
  \BibitemOpen
  \bibfield  {author} {\bibinfo {author} {\bibfnamefont {M.-K.}\ \bibnamefont
  {Lee}}\ and\ \bibinfo {author} {\bibfnamefont {M.}~\bibnamefont
  {Mochizuki}},\ }\bibfield  {title} {\bibinfo {title} {{Handwritten digit
  recognition by spin waves in a Skyrmion reservoir}},\ }\href
  {https://doi.org/10.1038/s41598-023-46677-w} {\bibfield  {journal} {\bibinfo
  {journal} {Scientific Reports}\ }\textbf {\bibinfo {volume} {13}},\ \bibinfo
  {pages} {19423} (\bibinfo {year} {2023})}\BibitemShut {NoStop}%
\bibitem [{\citenamefont {Bechler}\ and\ \citenamefont
  {Masell}(2023)}]{Bechler2023}%
  \BibitemOpen
  \bibfield  {author} {\bibinfo {author} {\bibfnamefont {N.}~\bibnamefont
  {Bechler}}\ and\ \bibinfo {author} {\bibfnamefont {J.}~\bibnamefont
  {Masell}},\ }\bibfield  {title} {\bibinfo {title} {Helitronics as a potential
  building block for classical and unconventional computing},\ }\href
  {https://doi.org/10.1088/2634-4386/ace549} {\bibfield  {journal} {\bibinfo
  {journal} {Neuromorphic Computing and Engineering}\ }\textbf {\bibinfo
  {volume} {3}},\ \bibinfo {pages} {034003} (\bibinfo {year}
  {2023})}\BibitemShut {NoStop}%
\bibitem [{\citenamefont {Lee}\ \emph {et~al.}(2023{\natexlab{b}})\citenamefont
  {Lee}, \citenamefont {Wei}, \citenamefont {Stenning}, \citenamefont
  {Gartside}, \citenamefont {Prestwood}, \citenamefont {Seki}, \citenamefont
  {Aqeel}, \citenamefont {Karube}, \citenamefont {Kanazawa}, \citenamefont
  {Taguchi}, \citenamefont {Back}, \citenamefont {Tokura}, \citenamefont
  {Branford},\ and\ \citenamefont {Kurebayashi}}]{Lee2023b}%
  \BibitemOpen
  \bibfield  {author} {\bibinfo {author} {\bibfnamefont {O.}~\bibnamefont
  {Lee}}, \bibinfo {author} {\bibfnamefont {T.}~\bibnamefont {Wei}}, \bibinfo
  {author} {\bibfnamefont {K.~D.}\ \bibnamefont {Stenning}}, \bibinfo {author}
  {\bibfnamefont {J.~C.}\ \bibnamefont {Gartside}}, \bibinfo {author}
  {\bibfnamefont {D.}~\bibnamefont {Prestwood}}, \bibinfo {author}
  {\bibfnamefont {S.}~\bibnamefont {Seki}}, \bibinfo {author} {\bibfnamefont
  {A.}~\bibnamefont {Aqeel}}, \bibinfo {author} {\bibfnamefont
  {K.}~\bibnamefont {Karube}}, \bibinfo {author} {\bibfnamefont
  {N.}~\bibnamefont {Kanazawa}}, \bibinfo {author} {\bibfnamefont
  {Y.}~\bibnamefont {Taguchi}}, \bibinfo {author} {\bibfnamefont
  {C.}~\bibnamefont {Back}}, \bibinfo {author} {\bibfnamefont {Y.}~\bibnamefont
  {Tokura}}, \bibinfo {author} {\bibfnamefont {W.~R.}\ \bibnamefont
  {Branford}},\ and\ \bibinfo {author} {\bibfnamefont {H.}~\bibnamefont
  {Kurebayashi}},\ }\bibfield  {title} {\bibinfo {title} {Task-adaptive
  physical reservoir computing},\ }\href
  {https://doi.org/10.1038/s41563-023-01698-8} {\bibfield  {journal} {\bibinfo
  {journal} {Nature Materials}\ ,\ \bibinfo {pages} {1}} (\bibinfo {year}
  {2023}{\natexlab{b}})}\BibitemShut {NoStop}%
\bibitem [{\citenamefont {Msiska}\ \emph {et~al.}(2023)\citenamefont {Msiska},
  \citenamefont {Love}, \citenamefont {Mulkers}, \citenamefont {Leliaert},\
  and\ \citenamefont {Everschor-Sitte}}]{Msiska2023}%
  \BibitemOpen
  \bibfield  {author} {\bibinfo {author} {\bibfnamefont {R.}~\bibnamefont
  {Msiska}}, \bibinfo {author} {\bibfnamefont {J.}~\bibnamefont {Love}},
  \bibinfo {author} {\bibfnamefont {J.}~\bibnamefont {Mulkers}}, \bibinfo
  {author} {\bibfnamefont {J.}~\bibnamefont {Leliaert}},\ and\ \bibinfo
  {author} {\bibfnamefont {K.}~\bibnamefont {Everschor-Sitte}},\ }\bibfield
  {title} {\bibinfo {title} {{Audio Classification with Skyrmion Reservoirs}},\
  }\href {https://doi.org/https://doi.org/10.1002/aisy.202200388} {\bibfield
  {journal} {\bibinfo  {journal} {Advanced Intelligent Systems}\ }\textbf
  {\bibinfo {volume} {2023}},\ \bibinfo {pages} {2200388} (\bibinfo {year}
  {2023})}\BibitemShut {NoStop}%
\bibitem [{\citenamefont {Chen}\ \emph {et~al.}(2023)\citenamefont {Chen},
  \citenamefont {Li}, \citenamefont {Fan}, \citenamefont {Dong}, \citenamefont
  {Chen}, \citenamefont {Qin}, \citenamefont {Zeng}, \citenamefont {Lu},
  \citenamefont {Zhou}, \citenamefont {Gao},\ and\ \citenamefont
  {Liu}}]{Chen2023}%
  \BibitemOpen
  \bibfield  {author} {\bibinfo {author} {\bibfnamefont {Z.}~\bibnamefont
  {Chen}}, \bibinfo {author} {\bibfnamefont {W.}~\bibnamefont {Li}}, \bibinfo
  {author} {\bibfnamefont {Z.}~\bibnamefont {Fan}}, \bibinfo {author}
  {\bibfnamefont {S.}~\bibnamefont {Dong}}, \bibinfo {author} {\bibfnamefont
  {Y.}~\bibnamefont {Chen}}, \bibinfo {author} {\bibfnamefont {M.}~\bibnamefont
  {Qin}}, \bibinfo {author} {\bibfnamefont {M.}~\bibnamefont {Zeng}}, \bibinfo
  {author} {\bibfnamefont {X.}~\bibnamefont {Lu}}, \bibinfo {author}
  {\bibfnamefont {G.}~\bibnamefont {Zhou}}, \bibinfo {author} {\bibfnamefont
  {X.}~\bibnamefont {Gao}},\ and\ \bibinfo {author} {\bibfnamefont {J.-M.}\
  \bibnamefont {Liu}},\ }\bibfield  {title} {\bibinfo {title}
  {All-ferroelectric implementation of reservoir computing},\ }\href
  {https://doi.org/https://doi.org/10.1038/s41467-023-39371-y} {\bibfield
  {journal} {\bibinfo  {journal} {Nature Communications}\ }\textbf {\bibinfo
  {volume} {14}} (\bibinfo {year} {2023})}\BibitemShut {NoStop}%
\bibitem [{\citenamefont {Tagantsev}\ \emph {et~al.}(2010)\citenamefont
  {Tagantsev}, \citenamefont {Cross},\ and\ \citenamefont
  {Fousek}}]{Tagantsev2010}%
  \BibitemOpen
  \bibfield  {author} {\bibinfo {author} {\bibfnamefont {A.~K.}\ \bibnamefont
  {Tagantsev}}, \bibinfo {author} {\bibfnamefont {L.~E.}\ \bibnamefont
  {Cross}},\ and\ \bibinfo {author} {\bibfnamefont {J.}~\bibnamefont
  {Fousek}},\ }\href {https://doi.org/10.1007/978-1-4419-1417-0} {\emph
  {\bibinfo {title} {Domains in ferroic crystals and thin films}}},\
  Vol.~\bibinfo {volume} {13}\ (\bibinfo  {publisher} {Springer},\ \bibinfo
  {year} {2010})\BibitemShut {NoStop}%
\bibitem [{\citenamefont {Yang}\ and\ \citenamefont {Alexe}(2018)}]{Yang2018}%
  \BibitemOpen
  \bibfield  {author} {\bibinfo {author} {\bibfnamefont {M.-M.}\ \bibnamefont
  {Yang}}\ and\ \bibinfo {author} {\bibfnamefont {M.}~\bibnamefont {Alexe}},\
  }\bibfield  {title} {\bibinfo {title} {{Light-induced reversible control of
  ferroelectric polarization in BiFeO3}},\ }\href
  {https://doi.org/10.1002/adma.201704908} {\bibfield  {journal} {\bibinfo
  {journal} {Advanced Materials}\ }\textbf {\bibinfo {volume} {30}},\ \bibinfo
  {pages} {1704908} (\bibinfo {year} {2018})}\BibitemShut {NoStop}%
\bibitem [{\citenamefont {Lu}\ \emph {et~al.}(2012)\citenamefont {Lu},
  \citenamefont {Bark}, \citenamefont {Esque De Los~Ojos}, \citenamefont
  {Alcala}, \citenamefont {Eom}, \citenamefont {Catalan},\ and\ \citenamefont
  {Gruverman}}]{Lu2012}%
  \BibitemOpen
  \bibfield  {author} {\bibinfo {author} {\bibfnamefont {H.}~\bibnamefont
  {Lu}}, \bibinfo {author} {\bibfnamefont {C.-W.}\ \bibnamefont {Bark}},
  \bibinfo {author} {\bibfnamefont {D.}~\bibnamefont {Esque De Los~Ojos}},
  \bibinfo {author} {\bibfnamefont {J.}~\bibnamefont {Alcala}}, \bibinfo
  {author} {\bibfnamefont {C.-B.}\ \bibnamefont {Eom}}, \bibinfo {author}
  {\bibfnamefont {G.}~\bibnamefont {Catalan}},\ and\ \bibinfo {author}
  {\bibfnamefont {A.}~\bibnamefont {Gruverman}},\ }\bibfield  {title} {\bibinfo
  {title} {Mechanical writing of ferroelectric polarization},\ }\href
  {https://doi.org/10.1126/science.1218693} {\bibfield  {journal} {\bibinfo
  {journal} {Science}\ }\textbf {\bibinfo {volume} {336}},\ \bibinfo {pages}
  {59} (\bibinfo {year} {2012})}\BibitemShut {NoStop}%
\bibitem [{\citenamefont {Kim}\ \emph {et~al.}(2023)\citenamefont {Kim},
  \citenamefont {Kim}, \citenamefont {Yun}, \citenamefont {Lee}, \citenamefont
  {Seo},\ and\ \citenamefont {Kim}}]{Kim2023}%
  \BibitemOpen
  \bibfield  {author} {\bibinfo {author} {\bibfnamefont {D.}~\bibnamefont
  {Kim}}, \bibinfo {author} {\bibfnamefont {J.}~\bibnamefont {Kim}}, \bibinfo
  {author} {\bibfnamefont {S.}~\bibnamefont {Yun}}, \bibinfo {author}
  {\bibfnamefont {J.}~\bibnamefont {Lee}}, \bibinfo {author} {\bibfnamefont
  {E.}~\bibnamefont {Seo}},\ and\ \bibinfo {author} {\bibfnamefont
  {S.}~\bibnamefont {Kim}},\ }\bibfield  {title} {\bibinfo {title}
  {{Ferroelectric synaptic devices based on CMOS-compatible HfAlO{$_x$} for
  neuromorphic and reservoir computing applications}},\ }\href
  {https://doi.org/10.1039/d3nr01294h} {\bibfield  {journal} {\bibinfo
  {journal} {Nanoscale}\ }\textbf {\bibinfo {volume} {15}},\ \bibinfo {pages}
  {8366} (\bibinfo {year} {2023})}\BibitemShut {NoStop}%
\bibitem [{\citenamefont {Tang}\ \emph {et~al.}(2023)\citenamefont {Tang},
  \citenamefont {Mei}, \citenamefont {Zhan}, \citenamefont {Wang},
  \citenamefont {Chai}, \citenamefont {Xu}, \citenamefont {Wang}, \citenamefont
  {Wu},\ and\ \citenamefont {Chen}}]{Tang2023}%
  \BibitemOpen
  \bibfield  {author} {\bibinfo {author} {\bibfnamefont {M.}~\bibnamefont
  {Tang}}, \bibinfo {author} {\bibfnamefont {J.}~\bibnamefont {Mei}}, \bibinfo
  {author} {\bibfnamefont {X.}~\bibnamefont {Zhan}}, \bibinfo {author}
  {\bibfnamefont {C.}~\bibnamefont {Wang}}, \bibinfo {author} {\bibfnamefont
  {J.}~\bibnamefont {Chai}}, \bibinfo {author} {\bibfnamefont {H.}~\bibnamefont
  {Xu}}, \bibinfo {author} {\bibfnamefont {X.}~\bibnamefont {Wang}}, \bibinfo
  {author} {\bibfnamefont {J.}~\bibnamefont {Wu}},\ and\ \bibinfo {author}
  {\bibfnamefont {J.}~\bibnamefont {Chen}},\ }\bibfield  {title} {\bibinfo
  {title} {{Fully Ferroelectric-FETs Reservoir Computing Network for Temporal
  and Random Signal Processing}},\ }\href
  {https://doi.org/10.1109/ted.2023.3268152} {\bibfield  {journal} {\bibinfo
  {journal} {IEEE Transactions on Electron Devices}\ } (\bibinfo {year}
  {2023})}\BibitemShut {NoStop}%
\bibitem [{\citenamefont {Toprasertpong}\ \emph {et~al.}(2022)\citenamefont
  {Toprasertpong}, \citenamefont {Nako}, \citenamefont {Wang}, \citenamefont
  {Nakane}, \citenamefont {Takenaka},\ and\ \citenamefont
  {Takagi}}]{Toprasertpong2022}%
  \BibitemOpen
  \bibfield  {author} {\bibinfo {author} {\bibfnamefont {K.}~\bibnamefont
  {Toprasertpong}}, \bibinfo {author} {\bibfnamefont {E.}~\bibnamefont {Nako}},
  \bibinfo {author} {\bibfnamefont {Z.}~\bibnamefont {Wang}}, \bibinfo {author}
  {\bibfnamefont {R.}~\bibnamefont {Nakane}}, \bibinfo {author} {\bibfnamefont
  {M.}~\bibnamefont {Takenaka}},\ and\ \bibinfo {author} {\bibfnamefont
  {S.}~\bibnamefont {Takagi}},\ }\bibfield  {title} {\bibinfo {title}
  {Reservoir computing on a silicon platform with a ferroelectric field-effect
  transistor},\ }\href
  {https://doi.org/https://doi.org/10.1038/s44172-022-00021-8} {\bibfield
  {journal} {\bibinfo  {journal} {Communications Engineering}\ }\textbf
  {\bibinfo {volume} {1}},\ \bibinfo {pages} {21} (\bibinfo {year}
  {2022})}\BibitemShut {NoStop}%
\bibitem [{\citenamefont {Falcone}\ \emph {et~al.}(2022)\citenamefont
  {Falcone}, \citenamefont {Halter}, \citenamefont {B{\'e}gon-Lours},\ and\
  \citenamefont {Offrein}}]{Falcone2022}%
  \BibitemOpen
  \bibfield  {author} {\bibinfo {author} {\bibfnamefont {D.~F.}\ \bibnamefont
  {Falcone}}, \bibinfo {author} {\bibfnamefont {M.}~\bibnamefont {Halter}},
  \bibinfo {author} {\bibfnamefont {L.}~\bibnamefont {B{\'e}gon-Lours}},\ and\
  \bibinfo {author} {\bibfnamefont {B.~J.}\ \bibnamefont {Offrein}},\
  }\bibfield  {title} {\bibinfo {title} {{Back-End, CMOS-Compatible
  Ferroelectric FinFET for Synaptic Weights}},\ }\href
  {https://doi.org/https://doi.org/10.3389/femat.2022.849879} {\bibfield
  {journal} {\bibinfo  {journal} {Frontiers in Electronic Materials}\ }\textbf
  {\bibinfo {volume} {2}},\ \bibinfo {pages} {849879} (\bibinfo {year}
  {2022})}\BibitemShut {NoStop}%
\bibitem [{\citenamefont {Eliseev}\ \emph {et~al.}(2011)\citenamefont
  {Eliseev}, \citenamefont {Morozovska}, \citenamefont {Svechnikov},
  \citenamefont {Gopalan},\ and\ \citenamefont {Shur}}]{Eliseev2011}%
  \BibitemOpen
  \bibfield  {author} {\bibinfo {author} {\bibfnamefont {E.}~\bibnamefont
  {Eliseev}}, \bibinfo {author} {\bibfnamefont {A.}~\bibnamefont {Morozovska}},
  \bibinfo {author} {\bibfnamefont {G.}~\bibnamefont {Svechnikov}}, \bibinfo
  {author} {\bibfnamefont {V.}~\bibnamefont {Gopalan}},\ and\ \bibinfo {author}
  {\bibfnamefont {V.~Y.}\ \bibnamefont {Shur}},\ }\bibfield  {title} {\bibinfo
  {title} {Static conductivity of charged domain walls in uniaxial
  ferroelectric semiconductors},\ }\href
  {https://doi.org/10.1103/PhysRevB.83.235313} {\bibfield  {journal} {\bibinfo
  {journal} {Physical Review B}\ }\textbf {\bibinfo {volume} {83}},\ \bibinfo
  {pages} {235313} (\bibinfo {year} {2011})}\BibitemShut {NoStop}%
\bibitem [{\citenamefont {Meier}\ \emph {et~al.}(2012)\citenamefont {Meier},
  \citenamefont {Seidel}, \citenamefont {Cano}, \citenamefont {Delaney},
  \citenamefont {Kumagai}, \citenamefont {Mostovoy}, \citenamefont {Spaldin},
  \citenamefont {Ramesh},\ and\ \citenamefont {Fiebig}}]{Meier2012}%
  \BibitemOpen
  \bibfield  {author} {\bibinfo {author} {\bibfnamefont {D.}~\bibnamefont
  {Meier}}, \bibinfo {author} {\bibfnamefont {J.}~\bibnamefont {Seidel}},
  \bibinfo {author} {\bibfnamefont {A.}~\bibnamefont {Cano}}, \bibinfo {author}
  {\bibfnamefont {K.}~\bibnamefont {Delaney}}, \bibinfo {author} {\bibfnamefont
  {Y.}~\bibnamefont {Kumagai}}, \bibinfo {author} {\bibfnamefont
  {M.}~\bibnamefont {Mostovoy}}, \bibinfo {author} {\bibfnamefont {N.~A.}\
  \bibnamefont {Spaldin}}, \bibinfo {author} {\bibfnamefont {R.}~\bibnamefont
  {Ramesh}},\ and\ \bibinfo {author} {\bibfnamefont {M.}~\bibnamefont
  {Fiebig}},\ }\bibfield  {title} {\bibinfo {title} {Anisotropic conductance at
  improper ferroelectric domain walls},\ }\href
  {https://doi.org/10.1038/nmat3249} {\bibfield  {journal} {\bibinfo  {journal}
  {Nature Materials}\ }\textbf {\bibinfo {volume} {11}},\ \bibinfo {pages}
  {284} (\bibinfo {year} {2012})}\BibitemShut {NoStop}%
\bibitem [{\citenamefont {Roede}\ \emph {et~al.}(2021)\citenamefont {Roede},
  \citenamefont {Mosberg}, \citenamefont {Evans}, \citenamefont {Bourret},
  \citenamefont {Yan}, \citenamefont {van Helvoort},\ and\ \citenamefont
  {Meier}}]{Roede2021}%
  \BibitemOpen
  \bibfield  {author} {\bibinfo {author} {\bibfnamefont {E.~D.}\ \bibnamefont
  {Roede}}, \bibinfo {author} {\bibfnamefont {A.~B.}\ \bibnamefont {Mosberg}},
  \bibinfo {author} {\bibfnamefont {D.~M.}\ \bibnamefont {Evans}}, \bibinfo
  {author} {\bibfnamefont {E.}~\bibnamefont {Bourret}}, \bibinfo {author}
  {\bibfnamefont {Z.}~\bibnamefont {Yan}}, \bibinfo {author} {\bibfnamefont
  {A.~T.~J.}\ \bibnamefont {van Helvoort}},\ and\ \bibinfo {author}
  {\bibfnamefont {D.}~\bibnamefont {Meier}},\ }\bibfield  {title} {\bibinfo
  {title} {Contact-free reversible switching of improper ferroelectric domains
  by electron and ion irradiation},\ }\href
  {https://doi.org/https://doi.org/10.1063/5.0038909} {\bibfield  {journal}
  {\bibinfo  {journal} {APL Materials}\ }\textbf {\bibinfo {volume} {9}},\
  \bibinfo {pages} {021105} (\bibinfo {year} {2021})}\BibitemShut {NoStop}%
\bibitem [{\citenamefont {Jiang}\ \emph {et~al.}(2018)\citenamefont {Jiang},
  \citenamefont {Bai}, \citenamefont {Chen}, \citenamefont {He}, \citenamefont
  {Zhang}, \citenamefont {Zhang}, \citenamefont {Shi}, \citenamefont {Park},
  \citenamefont {Scott}, \citenamefont {Hwang},\ and\ \citenamefont
  {Jiang}}]{Jiang2018}%
  \BibitemOpen
  \bibfield  {author} {\bibinfo {author} {\bibfnamefont {J.}~\bibnamefont
  {Jiang}}, \bibinfo {author} {\bibfnamefont {Z.~L.}\ \bibnamefont {Bai}},
  \bibinfo {author} {\bibfnamefont {Z.~H.}\ \bibnamefont {Chen}}, \bibinfo
  {author} {\bibfnamefont {L.}~\bibnamefont {He}}, \bibinfo {author}
  {\bibfnamefont {D.~W.}\ \bibnamefont {Zhang}}, \bibinfo {author}
  {\bibfnamefont {Q.~H.}\ \bibnamefont {Zhang}}, \bibinfo {author}
  {\bibfnamefont {J.~A.}\ \bibnamefont {Shi}}, \bibinfo {author} {\bibfnamefont
  {M.~H.}\ \bibnamefont {Park}}, \bibinfo {author} {\bibfnamefont {J.~F.}\
  \bibnamefont {Scott}}, \bibinfo {author} {\bibfnamefont {C.~S.}\ \bibnamefont
  {Hwang}},\ and\ \bibinfo {author} {\bibfnamefont {A.~Q.}\ \bibnamefont
  {Jiang}},\ }\bibfield  {title} {\bibinfo {title} {Temporary formation of
  highly conducting domain walls for non-destructive read-out of ferroelectric
  domain-wall resistance switching memories},\ }\href
  {https://doi.org/10.1038/nmat5028} {\bibfield  {journal} {\bibinfo  {journal}
  {Nature materials}\ }\textbf {\bibinfo {volume} {17}},\ \bibinfo {pages} {49}
  (\bibinfo {year} {2018})}\BibitemShut {NoStop}%
\bibitem [{\citenamefont {Choi}\ \emph {et~al.}(2010)\citenamefont {Choi},
  \citenamefont {Horibe}, \citenamefont {Yi}, \citenamefont {Choi},
  \citenamefont {Wu},\ and\ \citenamefont {Cheong}}]{Choi2010}%
  \BibitemOpen
  \bibfield  {author} {\bibinfo {author} {\bibfnamefont {T.}~\bibnamefont
  {Choi}}, \bibinfo {author} {\bibfnamefont {Y.}~\bibnamefont {Horibe}},
  \bibinfo {author} {\bibfnamefont {H.}~\bibnamefont {Yi}}, \bibinfo {author}
  {\bibfnamefont {Y.~J.}\ \bibnamefont {Choi}}, \bibinfo {author}
  {\bibfnamefont {W.}~\bibnamefont {Wu}},\ and\ \bibinfo {author}
  {\bibfnamefont {S.-W.}\ \bibnamefont {Cheong}},\ }\bibfield  {title}
  {\bibinfo {title} {{Insulating interlocked ferroelectric and structural
  antiphase domain walls in multiferroic YMnO3}},\ }\href
  {https://doi.org/10.1038/nmat2632} {\bibfield  {journal} {\bibinfo  {journal}
  {Nature materials}\ }\textbf {\bibinfo {volume} {9}},\ \bibinfo {pages} {253}
  (\bibinfo {year} {2010})}\BibitemShut {NoStop}%
\bibitem [{\citenamefont {Han}\ \emph {et~al.}(2013)\citenamefont {Han},
  \citenamefont {Zhu}, \citenamefont {Wu}, \citenamefont {Aoki}, \citenamefont
  {Volkov}, \citenamefont {Wang}, \citenamefont {Chae}, \citenamefont {Oh},\
  and\ \citenamefont {Cheong}}]{Han2013}%
  \BibitemOpen
  \bibfield  {author} {\bibinfo {author} {\bibfnamefont {M.-G.}\ \bibnamefont
  {Han}}, \bibinfo {author} {\bibfnamefont {Y.}~\bibnamefont {Zhu}}, \bibinfo
  {author} {\bibfnamefont {L.}~\bibnamefont {Wu}}, \bibinfo {author}
  {\bibfnamefont {T.}~\bibnamefont {Aoki}}, \bibinfo {author} {\bibfnamefont
  {V.}~\bibnamefont {Volkov}}, \bibinfo {author} {\bibfnamefont
  {X.}~\bibnamefont {Wang}}, \bibinfo {author} {\bibfnamefont {S.~C.}\
  \bibnamefont {Chae}}, \bibinfo {author} {\bibfnamefont {Y.~S.}\ \bibnamefont
  {Oh}},\ and\ \bibinfo {author} {\bibfnamefont {S.-W.}\ \bibnamefont
  {Cheong}},\ }\bibfield  {title} {\bibinfo {title} {Ferroelectric switching
  dynamics of topological vortex domains in a hexagonal manganite},\ }\href
  {https://doi.org/https://doi.org/10.1002/adma.201204766} {\bibfield
  {journal} {\bibinfo  {journal} {Advanced Materials}\ }\textbf {\bibinfo
  {volume} {25}},\ \bibinfo {pages} {2415} (\bibinfo {year}
  {2013})}\BibitemShut {NoStop}%
\bibitem [{\citenamefont {Grigoriev}\ \emph {et~al.}(2006)\citenamefont
  {Grigoriev}, \citenamefont {Do}, \citenamefont {Kim}, \citenamefont {Eom},
  \citenamefont {Adams}, \citenamefont {Dufresne},\ and\ \citenamefont
  {Evans}}]{Grigoriev2006}%
  \BibitemOpen
  \bibfield  {author} {\bibinfo {author} {\bibfnamefont {A.}~\bibnamefont
  {Grigoriev}}, \bibinfo {author} {\bibfnamefont {D.-H.}\ \bibnamefont {Do}},
  \bibinfo {author} {\bibfnamefont {D.~M.}\ \bibnamefont {Kim}}, \bibinfo
  {author} {\bibfnamefont {C.-B.}\ \bibnamefont {Eom}}, \bibinfo {author}
  {\bibfnamefont {B.}~\bibnamefont {Adams}}, \bibinfo {author} {\bibfnamefont
  {E.~M.}\ \bibnamefont {Dufresne}},\ and\ \bibinfo {author} {\bibfnamefont
  {P.~G.}\ \bibnamefont {Evans}},\ }\bibfield  {title} {\bibinfo {title}
  {{Nanosecond domain wall dynamics in ferroelectric Pb(Zr,Ti)O3 thin films}},\
  }\href {https://doi.org/10.1103/PhysRevLett.96.187601} {\bibfield  {journal}
  {\bibinfo  {journal} {Physical Review Letters}\ }\textbf {\bibinfo {volume}
  {96}},\ \bibinfo {pages} {187601} (\bibinfo {year} {2006})}\BibitemShut
  {NoStop}%
\bibitem [{\citenamefont {Holtz}\ \emph {et~al.}(2017)\citenamefont {Holtz},
  \citenamefont {Shapovalov}, \citenamefont {Mundy}, \citenamefont {Chang},
  \citenamefont {Yan}, \citenamefont {Bourret}, \citenamefont {Muller},
  \citenamefont {Meier},\ and\ \citenamefont {Cano}}]{Holtz2017}%
  \BibitemOpen
  \bibfield  {author} {\bibinfo {author} {\bibfnamefont {M.~E.}\ \bibnamefont
  {Holtz}}, \bibinfo {author} {\bibfnamefont {K.}~\bibnamefont {Shapovalov}},
  \bibinfo {author} {\bibfnamefont {J.~A.}\ \bibnamefont {Mundy}}, \bibinfo
  {author} {\bibfnamefont {C.~S.}\ \bibnamefont {Chang}}, \bibinfo {author}
  {\bibfnamefont {Z.}~\bibnamefont {Yan}}, \bibinfo {author} {\bibfnamefont
  {E.}~\bibnamefont {Bourret}}, \bibinfo {author} {\bibfnamefont {D.~A.}\
  \bibnamefont {Muller}}, \bibinfo {author} {\bibfnamefont {D.}~\bibnamefont
  {Meier}},\ and\ \bibinfo {author} {\bibfnamefont {A.}~\bibnamefont {Cano}},\
  }\bibfield  {title} {\bibinfo {title} {Topological defects in hexagonal
  manganites: Inner structure and emergent electrostatics},\ }\href
  {https://doi.org/10.1021/acs.nanolett.7b01288} {\bibfield  {journal}
  {\bibinfo  {journal} {Nano Letters}\ }\textbf {\bibinfo {volume} {17}},\
  \bibinfo {pages} {5883} (\bibinfo {year} {2017})}\BibitemShut {NoStop}%
\bibitem [{\citenamefont {Wang}\ \emph {et~al.}(2020)\citenamefont {Wang},
  \citenamefont {Feng}, \citenamefont {Zhu}, \citenamefont {Tang},
  \citenamefont {Yang}, \citenamefont {Zou}, \citenamefont {Geng},
  \citenamefont {Han}, \citenamefont {Guo}, \citenamefont {Wu},\ and\
  \citenamefont {Ma}}]{Wang2020}%
  \BibitemOpen
  \bibfield  {author} {\bibinfo {author} {\bibfnamefont {Y.}~\bibnamefont
  {Wang}}, \bibinfo {author} {\bibfnamefont {Y.}~\bibnamefont {Feng}}, \bibinfo
  {author} {\bibfnamefont {Y.}~\bibnamefont {Zhu}}, \bibinfo {author}
  {\bibfnamefont {Y.}~\bibnamefont {Tang}}, \bibinfo {author} {\bibfnamefont
  {L.}~\bibnamefont {Yang}}, \bibinfo {author} {\bibfnamefont {M.}~\bibnamefont
  {Zou}}, \bibinfo {author} {\bibfnamefont {W.}~\bibnamefont {Geng}}, \bibinfo
  {author} {\bibfnamefont {M.}~\bibnamefont {Han}}, \bibinfo {author}
  {\bibfnamefont {X.}~\bibnamefont {Guo}}, \bibinfo {author} {\bibfnamefont
  {B.}~\bibnamefont {Wu}},\ and\ \bibinfo {author} {\bibfnamefont
  {X.}~\bibnamefont {Ma}},\ }\bibfield  {title} {\bibinfo {title} {Polar meron
  lattice in strained oxide ferroelectrics},\ }\href
  {https://doi.org/10.1038/s41563-020-0694-8} {\bibfield  {journal} {\bibinfo
  {journal} {Nature Materials}\ }\textbf {\bibinfo {volume} {19}},\ \bibinfo
  {pages} {881} (\bibinfo {year} {2020})}\BibitemShut {NoStop}%
\bibitem [{\citenamefont {Luk'yanchuk}\ \emph {et~al.}(2020)\citenamefont
  {Luk'yanchuk}, \citenamefont {Tikhonov}, \citenamefont {Razumnaya},\ and\
  \citenamefont {Vinokur}}]{Luk2020}%
  \BibitemOpen
  \bibfield  {author} {\bibinfo {author} {\bibfnamefont {I.}~\bibnamefont
  {Luk'yanchuk}}, \bibinfo {author} {\bibfnamefont {Y.}~\bibnamefont
  {Tikhonov}}, \bibinfo {author} {\bibfnamefont {A.}~\bibnamefont
  {Razumnaya}},\ and\ \bibinfo {author} {\bibfnamefont {V.}~\bibnamefont
  {Vinokur}},\ }\bibfield  {title} {\bibinfo {title} {Hopfions emerge in
  ferroelectrics},\ }\href {https://doi.org/10.1038/s41467-020-16258-w}
  {\bibfield  {journal} {\bibinfo  {journal} {Nature Communications}\ }\textbf
  {\bibinfo {volume} {11}},\ \bibinfo {pages} {2433} (\bibinfo {year}
  {2020})}\BibitemShut {NoStop}%
\bibitem [{\citenamefont {Jakob}\ \emph {et~al.}(2015)\citenamefont {Jakob},
  \citenamefont {Andres}, \citenamefont {Lilienblum}, \citenamefont {Yan},
  \citenamefont {Bourret}, \citenamefont {Ramesh}, \citenamefont {Fiebig},\
  and\ \citenamefont {Meier}}]{Schaab2015}%
  \BibitemOpen
  \bibfield  {author} {\bibinfo {author} {\bibfnamefont {S.}~\bibnamefont
  {Jakob}}, \bibinfo {author} {\bibfnamefont {C.}~\bibnamefont {Andres}},
  \bibinfo {author} {\bibfnamefont {M.}~\bibnamefont {Lilienblum}}, \bibinfo
  {author} {\bibfnamefont {Z.}~\bibnamefont {Yan}}, \bibinfo {author}
  {\bibfnamefont {E.}~\bibnamefont {Bourret}}, \bibinfo {author} {\bibfnamefont
  {R.}~\bibnamefont {Ramesh}}, \bibinfo {author} {\bibfnamefont
  {M.}~\bibnamefont {Fiebig}},\ and\ \bibinfo {author} {\bibfnamefont
  {D.}~\bibnamefont {Meier}},\ }\bibfield  {title} {\bibinfo {title}
  {Optimization of electronic domain-wall properties by aliovalent cation
  substitution},\ }\href {https://doi.org/10.1002/aelm.201500195} {\bibfield
  {journal} {\bibinfo  {journal} {Advanced Electronic Materials}\ }\textbf
  {\bibinfo {volume} {2}},\ \bibinfo {pages} {1500195} (\bibinfo {year}
  {2015})}\BibitemShut {NoStop}%
\bibitem [{\citenamefont {Fiebig}\ \emph {et~al.}(2016)\citenamefont {Fiebig},
  \citenamefont {Lottermoser}, \citenamefont {Meier},\ and\ \citenamefont
  {Trassin}}]{Fiebig2016}%
  \BibitemOpen
  \bibfield  {author} {\bibinfo {author} {\bibfnamefont {M.}~\bibnamefont
  {Fiebig}}, \bibinfo {author} {\bibfnamefont {T.}~\bibnamefont {Lottermoser}},
  \bibinfo {author} {\bibfnamefont {D.}~\bibnamefont {Meier}},\ and\ \bibinfo
  {author} {\bibfnamefont {M.}~\bibnamefont {Trassin}},\ }\bibfield  {title}
  {\bibinfo {title} {The evolution of multiferroics},\ }\href
  {https://doi.org/10.1038/natrevmats.2016.46} {\bibfield  {journal} {\bibinfo
  {journal} {Nature Reviews Materials}\ }\textbf {\bibinfo {volume} {1}},\
  \bibinfo {pages} {16046} (\bibinfo {year} {2016})}\BibitemShut {NoStop}%
\bibitem [{\citenamefont {Nothhelfer}\ \emph {et~al.}(2022)\citenamefont
  {Nothhelfer}, \citenamefont {D\'{\i}az}, \citenamefont {Kessler},
  \citenamefont {Meng}, \citenamefont {Rizzi}, \citenamefont {Hals},\ and\
  \citenamefont {Everschor-Sitte}}]{Nothhelfer2022}%
  \BibitemOpen
  \bibfield  {author} {\bibinfo {author} {\bibfnamefont {J.}~\bibnamefont
  {Nothhelfer}}, \bibinfo {author} {\bibfnamefont {S.~A.}\ \bibnamefont
  {D\'{\i}az}}, \bibinfo {author} {\bibfnamefont {S.}~\bibnamefont {Kessler}},
  \bibinfo {author} {\bibfnamefont {T.}~\bibnamefont {Meng}}, \bibinfo {author}
  {\bibfnamefont {M.}~\bibnamefont {Rizzi}}, \bibinfo {author} {\bibfnamefont
  {K.~M.~D.}\ \bibnamefont {Hals}},\ and\ \bibinfo {author} {\bibfnamefont
  {K.}~\bibnamefont {Everschor-Sitte}},\ }\bibfield  {title} {\bibinfo {title}
  {{Steering Majorana braiding via skyrmion-vortex pairs: A scalable
  platform}},\ }\href {https://doi.org/10.1103/PhysRevB.105.224509} {\bibfield
  {journal} {\bibinfo  {journal} {Phys. Rev. B}\ }\textbf {\bibinfo {volume}
  {105}},\ \bibinfo {pages} {224509} (\bibinfo {year} {2022})}\BibitemShut
  {NoStop}%
\bibitem [{\citenamefont {Stenning}\ \emph {et~al.}(2022)\citenamefont
  {Stenning}, \citenamefont {Gartside}, \citenamefont {Manneschi},
  \citenamefont {Cheung}, \citenamefont {Chen}, \citenamefont {Vanstone},
  \citenamefont {Love}, \citenamefont {Holder}, \citenamefont {Caravelli},
  \citenamefont {Everschor-Sitte}, \citenamefont {Vasilaki},\ and\
  \citenamefont {Branford}}]{Stenning2022}%
  \BibitemOpen
  \bibfield  {author} {\bibinfo {author} {\bibfnamefont {K.~D.}\ \bibnamefont
  {Stenning}}, \bibinfo {author} {\bibfnamefont {J.~C.}\ \bibnamefont
  {Gartside}}, \bibinfo {author} {\bibfnamefont {L.}~\bibnamefont {Manneschi}},
  \bibinfo {author} {\bibfnamefont {C.~T.}\ \bibnamefont {Cheung}}, \bibinfo
  {author} {\bibfnamefont {T.}~\bibnamefont {Chen}}, \bibinfo {author}
  {\bibfnamefont {A.}~\bibnamefont {Vanstone}}, \bibinfo {author}
  {\bibfnamefont {J.}~\bibnamefont {Love}}, \bibinfo {author} {\bibfnamefont
  {H.~H.}\ \bibnamefont {Holder}}, \bibinfo {author} {\bibfnamefont
  {F.}~\bibnamefont {Caravelli}}, \bibinfo {author} {\bibfnamefont
  {K.}~\bibnamefont {Everschor-Sitte}}, \bibinfo {author} {\bibfnamefont
  {E.}~\bibnamefont {Vasilaki}},\ and\ \bibinfo {author} {\bibfnamefont
  {W.~R.}\ \bibnamefont {Branford}},\ }\bibfield  {title} {\bibinfo {title}
  {Adaptive programmable networks for in materia neuromorphic computing},\
  }\href {https://doi.org/10.48550/arXiv.2211.06373} {\bibfield  {journal}
  {\bibinfo  {journal} {arXiv:2211.06373}\ } (\bibinfo {year}
  {2022})}\BibitemShut {NoStop}%
\end{thebibliography}%
\end{document}